\documentclass{article}

\usepackage{PRIMEarxiv}

\usepackage[utf8]{inputenc} 
\usepackage[T1]{fontenc}    
\usepackage{hyperref}       
\usepackage{url}            
\usepackage{booktabs}       
\usepackage{amsfonts}       
\usepackage{nicefrac}       
\usepackage{microtype}      
\usepackage{lipsum}
\usepackage{fancyhdr}       
\usepackage{graphicx}       
\graphicspath{{media/}}     

\usepackage{siunitx}
\usepackage{dsfont}
\usepackage{subcaption}
\usepackage{lineno}
\usepackage{bm}
\usepackage{amsmath}
\usepackage{natbib}
\usepackage{authblk}

\usepackage{amssymb}
\usepackage{comment}
\usepackage{soul}
\usepackage{tabularx}
\usepackage{pdflscape}

\title{A unified neural background-error covariance model for midlatitude and tropical atmospheric data assimilation}

\author[1]{\textbf{Bo\v stjan Melinc}}
\author[1]{\textbf{Uro\v s Perkan}}
\author[2,1]{\textbf{\v{Z}iga Zaplotnik}}

\affil[1]{University of Ljubljana, Faculty of Mathematics and Physics, \protect\\Jadranska 19, 1000 Ljubljana, Slovenia\vspace{2mm}}
\affil[2]{European Centre for Medium-range Weather Forecasts, \protect\\Robert-Schuman-Platz 3, 53175 Bonn, Germany}

\date{} 

\usepackage{tikz}

\usepackage{atbegshi} 

\newcommand{\firstpagefooterpublished}{%
  \begin{tikzpicture}[remember picture, overlay]
    \node at ([yshift=1.5cm]current page.south) {
      \begin{minipage}{\textwidth}
        \centering
        \rule{\textwidth}{0.2pt}\\
        \textcolor{red}{\textbf{Peer-reviewed version available in JAMES}:} 
        \textcolor{blue}{\href{https://doi.org/10.1029/2025MS005360}{ https://doi.org/10.1029/2025MS005360}} 
      \end{minipage}
    };
  \end{tikzpicture}%
}

\AtBeginShipoutFirst{\firstpagefooterpublished}

\begin{document}
\maketitle

\begin{abstract}
Estimating background-error covariances remains a core challenge in variational data assimilation (DA). Operational systems typically approximate these covariances by transformations that separate geostrophically balanced components from unbalanced inertio-gravity modes -- an approach well-suited for the midlatitudes but less applicable in the tropics, where different physical balances prevail. This study estimates background-error covariances in a reduced-dimension latent space learned by a neural-network autoencoder (AE). The AE was trained using 40 years of ERA5 reanalysis data, enabling it to capture flow-dependent atmospheric balances from a diverse set of weather states.

We demonstrate that performing DA in the latent space yields analysis increments that preserve multivariate horizontal and vertical physical balances in both tropical and midlatitude atmosphere. Assimilating a single 500\,hPa geopotential height observation in the midlatitudes produces increments consistent with geostrophic and thermal wind balance, while assimilating a total column water vapor observation with a positive departure in the nearly-saturated tropical atmosphere generates an increment resembling the tropical response to (latent) heat-induced perturbations. The resulting increments are localized and flow-dependent, and shaped by orography and land-sea contrasts. 

Forecasts initialized from these analyses exhibit realistic weather evolution, including the excitation of an eastward-propagating Kelvin wave in the tropics. 
 
Finally, we explore the transition from using synthetic ensembles and a climatology-based background error covariance matrix to an operational ensemble of data assimilations (EDA). Despite significant compression-induced variance loss in some variables, latent-space assimilation produces balanced, flow-dependent increments -- highlighting its potential for ensemble-based latent-space 4D-Var.
\end{abstract}

\vspace{1cm}

\vspace{1cm}

\section{Introduction}

Global weather forecasting is an initial value problem. Given an estimate of the initial state of the atmosphere, the forecast model simulates its evolution~\citep{Kalnay2002AtmosphericPredictability}. 
The best estimate of the initial state, known as the \textit{analysis}, is obtained through data assimilation (DA). DA combines prior information from a previous short-range forecast (the \textit{background}) with new observations, accounting for their uncertainties and physical constraints.
One approach to this problem is Variational Data Assimilation ~\citep{Lahoz2010DAMakingSense, Park2009DAAtmosphereOceansHydrology}, which minimizes the cost function to estimate the analysis. When the background and observations are approximately concurrent, this leads to the three-dimensional variational (3D-Var) formulation. 

One of the key challenges in variational DA is the representation of background-error covariances~\citep{Bannister2008a}. These determine the relative weight of the background and observations and the spread of the information from observations in space and between variables. Analytically, the covariances are encapsulated in a $\mathbf{B}$-matrix, which includes the spatial covariances among all model variables.  
To avoid issues related to $\mathbf{B}$-matrix inversion, operational weather centers do not construct and store $\mathbf{B}$ but rather define it implicitly using control variable transform (CVT). The balancing terms in  CVTs, which map the control variables to model variables, are partly analytical and rely on the assumption that background errors can be adequately characterized by dividing them into vorticity-like components (quasi-balanced, associated with Rossby modes) and divergence-like components (quasi-unbalanced, associated with inertio-gravity modes,~\citet{Bannister2008b}).

While geostrophic balance dominates in the extratropical atmosphere, the governing physical balances in the tropics differ significantly due to a weaker Coriolis force. \citet{Matsuno1966Quasi} demonstrated that this leads to a rich spectrum of eigenmodes, fundamentally distinct to those in the extratropics. \citet{Zagar2004VariationalConstraint} therefore proposed to use balance relationships based on equatorial wave theory for tropical DA and demonstrated the effectiveness of this approach in a shallow water model (SWM) on an equatorial $\beta$-plane. Building on this, \citet{Koernich2008Combining} solved the 3D-Var cost function by separating the control variables associated with midlatitude modes and equatorial modes, determined based on the Hough modes (the eigenmodes of the linearised atmospheric motions) of certain equivalent depths and wavenumbers. This framework allowed the system to represent both tropical and extratropical covariances, though with some limitations, such as the incomplete balances. 
Despite their proven usefulness in diagnostic studies~\citep[e.g.,][]{Zagar2005BalancedErrors, Zagar2013BalanceEnsemble}, the normal modes (the three-dimensional generalization of Hough modes~\citep{Kasahara1981SpectralFunctions}) have not yet been implemented in the operational DA systems. 
Consequently, the operational weather centers still lack the representation of tropical balances in their systems \citep[e.g.,][]{ifs49r1DA}.

Recent years have seen a rapid increase in the use of machine learning (ML) to develop novel DA techniques (see~\citet{Cheng2023MachineReview}, \citet[Sec.~11.4]{Bach2024InverseApproach}, and \citet{Pasmans2025EnsemblePair} for comprehensive reviews). The idea of using ML for unifying the assimilation procedure in the midlatitudes and the tropics was first proposed by~\citet[hereafter MZ24]{Melinc20243D-VarAutoencoder}, who trained a variational autoencoder (VAE) for reproducing global 850\,hPa temperature (T850) fields from ERA5 reanalysis, and performed 3D-Var DA in its reduced-order latent space. Using single observation experiments, they showed that compressing the T850 into a latent space of 100 elements with a VAE produces a background-error covariance matrix $\mathbf{B}_z$ that effectively captures local covariances in both midlatitudes and tropics. Notably, although $\mathbf{B}_z$ is static and climatological, the background-error covariances in the physical space exhibited state-dependent features. \citet{Zheng2024GeneratingFramework} then applied the same approach to assimilate sea-surface temperature and demonstrated its feasibility with real-world observations, while \citet{Fan2025PhysicallySpace} performed 3D-Var and 4D-Var DA in the latent space of a vision transformer.
Their analysis increments in the midlatitudes followed the geostrophic balance
and were strongly affected by the background flow. Outside of latent-space DA, the physical consistency of analysis increments produced by the ML-based DA techniques received limited attention. \citet{Xu2025FuXi-DA:Observations} showed that in their system for assimilating satellite observations, an observation with a positive departure of temperature will lead to a physically-consistent decrease in the local relative humidity. Besides, \citet{Li2024FuXiEn4DVar:Constraints} showed physically meaningful analysis increments given the background flow and the time of the observation in their 4D-Var system. 


To date, ML-based methods for atmospheric DA were either only deterministic \citep[e.g.,][]{Xiao2023FengWu-4DVar:Assimilation, Chen2023Adas, Xiang2024ADAF, Sun2024FuXiWeather, Zheng2024GeneratingFramework, Fan2025PhysicallySpace, Xu2025FuXi-DA:Observations}, or used statistical or ML methods to directly estimate analysis uncertainty or to generate ensembles of analyses \citep[MZ24,][]{Li2024FuXiEn4DVar:Constraints, Chen2024FNP, Andry2025Appa:Assimilation}. However, none of them leveraged uncertainty information from ensembles of operational forecasts. 

In this paper, we present a unified background-error covariance model for data assimilation in both the tropical and midlatitude atmosphere. To achieve this, we extend the approach from MZ24 to a multilevel,  multivariate representation of the atmosphere. Section~\ref{sec:data methods} outlines the methodology for 3D-Var data assimilation in the latent space of a neural-network-based autoencoder (AE). 
Section~\ref{sec:singobs} presents a detailed evaluation of two single observation experiments: one in the midlatitudes and one in the tropics. We assess their consistency with expected physical balances and run forecasts initialized from the resulting analyses. In Section~\ref{sec:true EDA}, we estimate the flow-dependent background-error covariance model in the latent space by sampling an ensemble of backgrounds from operational ensemble of data assimilations (EDA) from ECMWF's Integrated Forecasting System (IFS). The discussion and conclusions are provided in Sec.~\ref{sec:discussion}.

\section{Data and methods}\label{sec:data methods}

Single observation 3D-Var data assimilation experiments and subsequent forecasts were performed using two neural networks with similar architecture and model variables: (1) a convolutional autoencoder (AE) for performing latent-space 3D-Var (Sec.~\ref{sec:3D-Var}), and (2) a convolutional U-Net, based on \citet{Perkan2025UsingModels}, used as a neural-network forecasting model to recursively predict the atmospheric states at 12-hour intervals.

\subsection{Data}

The NNs were trained on the ERA5 reanalysis dataset~\citep{Hersbach2020TheReanalysis}, retrieved from~\citet{CopernicusClimateChangeServiceClimateDataStore2023ERA5Present}. The training set consisted of hourly data from 1970 to 2014, the validation set covered 2015 to 2019, and the test set included 2020 to 2022. Data was downloaded on a regular $1^\circ\times1^\circ$ grid and regridded meridionally to avoid the singularities at the poles.

The NNs were trained to reconstruct or predict the variables listed in Table~\ref{tab:quantities}. Each physical variable was represented as a $180\times360$ field at each specified level and standardized by subtracting the climatological mean and dividing by the climatological standard deviation, computed individually for each grid point. The dataset included 20 dynamical variables -- geopotential height at four levels, zonal and meridional wind at five levels, temperature at four levels, mean sea level pressure, and total column water vapor -- and three static variables (land-sea mask, latitude, and surface elevation) which only served as inputs to the NNs, but were not part of their output.

\begin{table}[h]
    \caption{Fields used as inputs to the neural networks. The static fields at the bottom of the table (italicized) are excluded from the output. Surface temperature is a composite of soil temperature over land and sea surface temperature. The abbreviations are written in the same order as the levels.}
    \label{tab:quantities}
    \renewcommand{\arraystretch}{1.4} 
    \centering
    \begin{tabularx}{\linewidth}{l>{\raggedright\arraybackslash}X>{\raggedright\arraybackslash}X}
        \hline
        Quantity [unit] & Levels & Abbreviation  \\
        \hline
        Geopotential height [m] & 250\,hPa, 500\,hPa, 700\,hPa, 850\,hPa & Z250, Z500, Z700, Z850\\
        Zonal wind [m/s]  & 200\,hPa, 500\,hPa, 700\,hPa, 900\,hPa, 10\,m & U200, U500, U700, U900, U10m \\
        Meridional wind [m/s] &  200\,hPa, 500\,hPa, 700\,hPa, 900\,hPa, 10\,m & V200, V500, V700, V900, V10m \\
        Temperature [K]  &  500\,hPa, 850\,hPa, 2\,m, surface & T500, T850, T2m, ST \\
        Mean sea level pressure [hPa] &  - & MSLP \\
        Total column water vapor [kg/m$^2$] &  - & TCWV \\
        \textit{Land-sea mask} &  - & - \\
        \textit{Latitude} &  - & - \\
        \textit{Surface elevation} &  - & - \\
        \hline
    \end{tabularx}
\end{table}

\subsection{Neural networks}

The autoencoder (AE) consists of an encoder, which compresses the input into a latent representation, and a decoder, which reconstructs the original data from this compressed form. The standardized atmospheric state with shape $23 \times 180 \times 360$ is first passed through the encoder $E$ to obtain the \textit{latent vector} $\mathbf{z}$ containing 12100 elements. The latent vector is then input to the decoder $D$, which maps it back to the standardized physical space. The quality of the reconstruction is visualized in Fig.~\ref{fig:AE quality}. 

Similarly to the AE, the input fields are first compressed and then decompressed in a U-Net. Additionally, each encoder block is directly connected to its corresponding decoder block with the same field resolution via skip connections. These connections allow fine-scale spatial information to bypass the compression bottleneck, improving reconstruction accuracy, especially for localized features~\citep{Ronneberger2015UNet}. An example of the forecast quality is provided in Fig.~\ref{fig:NNfwd quality}. The technical details of the two applied neural networks are provided in Appendix~\ref{app:nn}.

\subsection{3D-Var in a latent space of an autoencoder}
\label{sec:3D-Var}
Three-dimensional variational data assimilation (3D-Var) estimates the most likely state of the atmosphere, the \textit{analysis}, by optimally combining information from the previous short-range forecast, the \textit{background}, with new observations. 
As shown in MZ24, the background state $\mathbf{x}_b$ can be represented in the AE's latent space as $\mathbf{z}_b = E \circ S(\mathbf{x}_b)$, and the analysis $\mathbf{z}_a$ is obtained by minimizing the cost function
\begin{eqnarray}
    \label{eq:J classic}
        J_z(\mathbf{z}) &= & \frac{1}{2}\big(\mathbf{z} - \mathbf{z}_b\big)^\top \mathbf{B}_z^{-1} \big(\mathbf{z} - \mathbf{z}_b\big) \nonumber \\  
    &&+ \  \frac{1}{2} \left\{\mathbf{y} - H\left[S^{-1} \circ D(\mathbf{z})\right]\right\}^\top \mathbf{R}^{-1} \left\{\mathbf{y} - H\left[S^{-1}\circ D(\mathbf{z})\right]\right\},
\end{eqnarray}
where $\mathbf{z}$ is the latent vector and $\mathbf{B}_z$ is the background-error covariance matrix in the AE's latent space. The vector $\mathbf{y}$ represents the observations, $H$ is the observation operator (in our case a bilinear interpolation), $D$ denotes the decoder and $E$ the encoder, $S$ is the standardization operator, $S^{-1}$ its inverse (destandardization), and $\mathbf{R}$ is the observation-error covariance matrix.

The derivation of the cost function $J_z$ assumes that background errors are Gaussian and unbiased. MZ24 avoided the potential issues with these assumptions by applying the cost function~\eqref{eq:J classic} in a latent space of a VAE, which ensured the Gaussian properties of the latent vector. 
In contrast, a ``standard'' autoencoder does not guarantee Gaussianity in its latent space. However, the latent vectors obtained by encoding the atmospheric states from the validation set exhibited near-Gaussian behavior, with a mean absolute skewness of latent elements of $0.12$ and mean absolute kurtosis of $0.19$. These values indicate that the latent space is sufficiently close to Gaussian for the latent-space 3D-Var formulation to remain valid.

We also assessed the bias in background $\mathbf{z}_b$ provided by the 24-hour forecast using the NN forecasting model. For each latent vector element, we compared the mean and standard deviation of the forecast error using forecasts initialized from the atmospheric states in the validation set at 1-hour intervals (Fig.~\ref{fig:B_and_bias}a). 
The background (and so its errors) can be considered unbiased since for most latent vector elements, the error standard deviation vastly exceeded the mean. In a few latent vector elements, the mean approached but did not exceed the standard deviation, indicating some bias, which we assumed negligible for this study.

We preconditioned the minimisation problem by taking $\boldsymbol{\chi} = \mathbf{L}_z^{-1} (\mathbf{z} - \mathbf{z}_b)$ and square-root preconditioner $\mathbf{L}_z = \mathbf{B}_z^{1/2}$ and rewrote the cost function~\eqref{eq:J classic} into
\begin{eqnarray}
    \label{eq:J preconditioned}
    J_\chi(\boldsymbol{\chi}) &=& \frac{1}{2}\,\boldsymbol{\chi}^\top \boldsymbol{\chi} \nonumber\\
    &&+ \  \frac{1}{2}\, \left\{\mathbf{y} - H\left[S^{-1}\circ D\left(\mathbf{z}_b + \mathbf{L}_z \boldsymbol{\chi}\right)\right]\right\}^\top \mathbf{R}^{-1} \left\{\mathbf{y} - H\left[S^{-1}\circ D\left(\mathbf{z}_b + \mathbf{L}_z \boldsymbol{\chi}\right)\right]\right\},
\end{eqnarray} 
Minimizing $\boldsymbol{\chi}$ instead of $\mathbf{z}$ reduces the condition number, leading to more stable minimisation and fewer iterations to achieve convergence~\citep{Bannister2008b}.
Further technical details on the minimization algorithm are provided in Appendix~\ref{app:minimization}.

\subsection{Modelling background-error covariances}
\label{sec:modelling B}
The climatological background-error covariance matrix for the experiments in Sec.~\ref{sec:singobs} was computed as
\begin{equation}
    \label{eq:Bz clim}
    \mathbf{B}^{clim}_z = \left\langle \left(\mathbf{z}_b - \mathbf{z}_t\right) \left(\mathbf{z}_b - \mathbf{z}_t\right)^\top\right\rangle,
\end{equation}
where $\mathbf{z}_b$ is the encoded 24-hour NN forecast that served as the background, $\mathbf{z}_t$ denotes the encoded ground ``truth'' (i.e., the encoded ERA5 reanalysis for a chosen date and time), and the angle brackets denote averaging of the outer product over the entire validation set. The details on the computation and properties of the background-error covariance matrix from the operational ECMWF IFS ensemble of data assimilations (EDA) are provided in Section~\ref{sec:true EDA}.

In MZ24, the $\mathbf{B}_z$-matrix was quasi-diagonal and retaining only its diagonal elements for inversion had no negative impact on the assimilation result. Similar findings were reported by \citet{Zheng2024GeneratingFramework} and \citet{Fan2025PhysicallySpace}. 
In our case, the diagonal elements of $\mathbf{B}_z$ were typically two orders of magnitude larger than the off-diagonals (Fig.~\ref{fig:B_and_bias}b). This allowed us to approximate $\mathbf{B}_z^{1/2}$ and $\mathbf{B}_z^{-1/2}$ using only the diagonal, which greatly simplified the computation.

\section{Single observation experiments}
\label{sec:singobs}
We present two single-observation experiments that illustrate the balance properties of the background-error covariance model in the midlatitudes and the tropics, and demonstrate their physical consistency. Each experiment applies a 24-hour forecast initialized from ERA5 reanalysis data on April 14, 2020, at 00\,UTC, which is then encoded into latent space. We generated 100-member ensemble of backgrounds by perturbing this encoded forecast using variances from the $\mathbf{B}_z$-matrix. Each perturbed background member has a corresponding perturbed observation set $\mathbf{y}$. This is obtained by first decoding and destandardizing the background ensemble members, computing their mean and then interpolating it to the observation location. We then added the preset observation departure $\mathbf{d}$ to that mean and perturbed the output according to the observation standard deviation. The procedure can be described as:
\begin{equation}
    \mathbf{y} = H \left[\left<S^{-1} \circ D (\mathbf{z}_b)\right>\right] + \mathbf{d} + \boldsymbol{\epsilon}_{o} \, ,
\end{equation}
where $\boldsymbol{\epsilon}_{o}\sim\mathcal{N}(\mathbf{0},\mathbf{R})$ denotes the random vector sampled from a zero-mean and diagonal observation-error covariance matrix, and angle brackets denote the ensemble mean.

Each of the 100 pairs of perturbed background and observations was used in a separate 3D-Var DA, resulting in 100 latent-space analyses. The resulting analysis increment $\delta\mathbf{x}_a$ was computed as the mean of the differences between the decoded analysis and the decoded background, i.e. 
\begin{equation}
    \delta\mathbf{x}_a = \left<S^{-1} \circ D (\mathbf{z}_a) - S^{-1} \circ D (\mathbf{z}_b)\right> .
\end{equation}

The notation in this section and Sec.~\ref{sec:true EDA} is organized as follows: the observation departure of a variable $\mathrm{V}$ is labeled as $d^\mathrm{V}$, the standard deviation of the observation is $\sigma^\mathrm{V}_o$, the analysis increment is $\delta^\mathrm{V}_a$, and the analysis standard deviation is $\sigma^\mathrm{V}_a$. 
The abbreviations are the same as in Table~\ref{tab:quantities}.

\subsection{Background-error covariance model in the midlatitudes -- geostrophic and thermal wind balance}

\label{sec:geostrophic adjustment}

In the midlatitudes, large-scale atmospheric flow above the planetary boundary layer tends to follow geostrophic balance. Assimilating an observation in this region should, in a well-tuned DA system, yield analysis increments that largely preserve this balance. To evaluate the physical realism of our DA system, we assimilated a single geopotential height observation at 500\,hPa (Z500), and studied the resulting analysis increment and the adjustment of the NN forecasting model to the induced perturbation.

\begin{figure}[h]
    \centering
    \includegraphics[width=\textwidth]{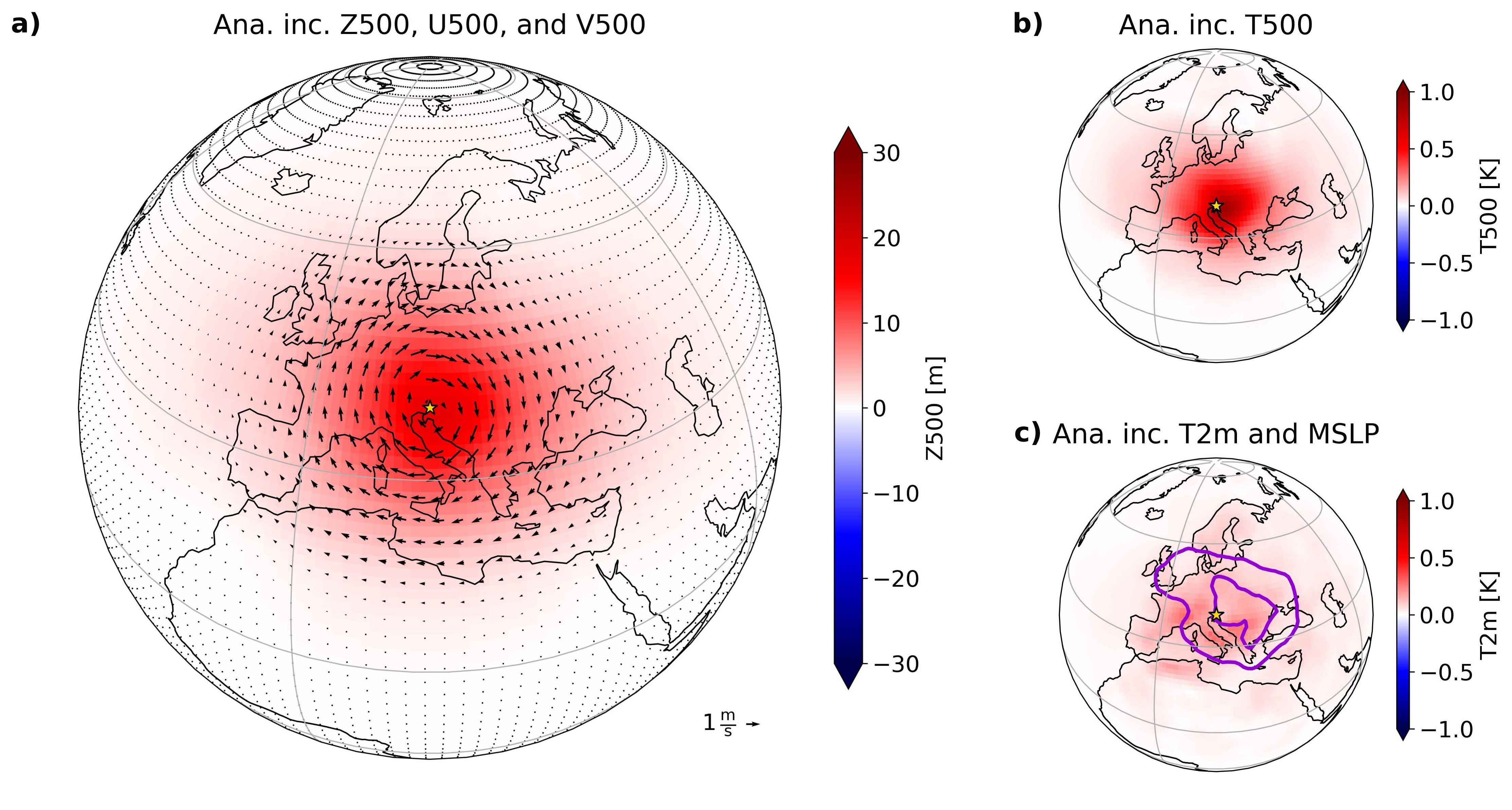}
    \caption{Analysis increments following an assimilation of Z500 observation above Ljubljana with departure of 30\,m and observation-error standard deviation of 10\,m. (a) Z500 increment (colors) and 500\,hPa horizontal wind increment (arrows); (b) T500 increment; (c) T2m increment (colors) and MSLP increment (the two purple contours denote  $+0.15$\,hPa, and $+0.30$\,hPa increments). The observation location is denoted by a golden star.}
    \label{fig:Ljubljana}
\end{figure}

We first simulated the Z500 observation above Ljubljana, Slovenia, ($46.1\,^\circ\mathrm{N}$, $14.5\,^\circ\mathrm{E}$) with departure of $d^\mathrm{Z500}=30$\,m and standard deviation of $\sigma_o^\mathrm{Z500}=10$\,m. Figure~\ref{fig:Ljubljana} shows the analysis increments for selected variables, with the full output provided in Fig.~\ref{fig:full Ljubljana}. At the 500\,hPa pressure level, where the observation was assimilated, the geostrophic balance is evident: a positive Z500 analysis increment centered on the observation location is accompanied by an anticyclonic wind increment (Fig.~\ref{fig:Ljubljana}a) and a positive temperature increment from the surface to the observation level (Fig.~\ref{fig:Ljubljana}b,c). The T500 increment is nearly isotropic, while the T2m increment's shape is affected by the land-sea distribution, with the larger increments over the land areas, where the background-error standard deviations are larger (Fig.~\ref{fig:full Ljubljana}, panel C7). An eastward tilt of temperature increments and a westward tilt of geopotential increments with height, in accordance with the quasi-geostrophic theory of developing Rossby waves, further proves the physical plausibility of the analysis increments (Fig.~\ref{fig:full Ljubljana}, row B).

A positive MSLP increment is located eastward of the maximum geopotential height increment. This is due to (1) negative (anticyclonic) relative vorticity advection and (2) large-scale convergence in the upper-troposphere on the eastern side of the Rossby wave ridge, which leads to an increase in the surface pressure below. A slight distortion of the increment is due to interaction with the orography (Alps, Dinaric Alps).

\begin{figure}[h!]
    \centering
    \includegraphics[width=\textwidth]{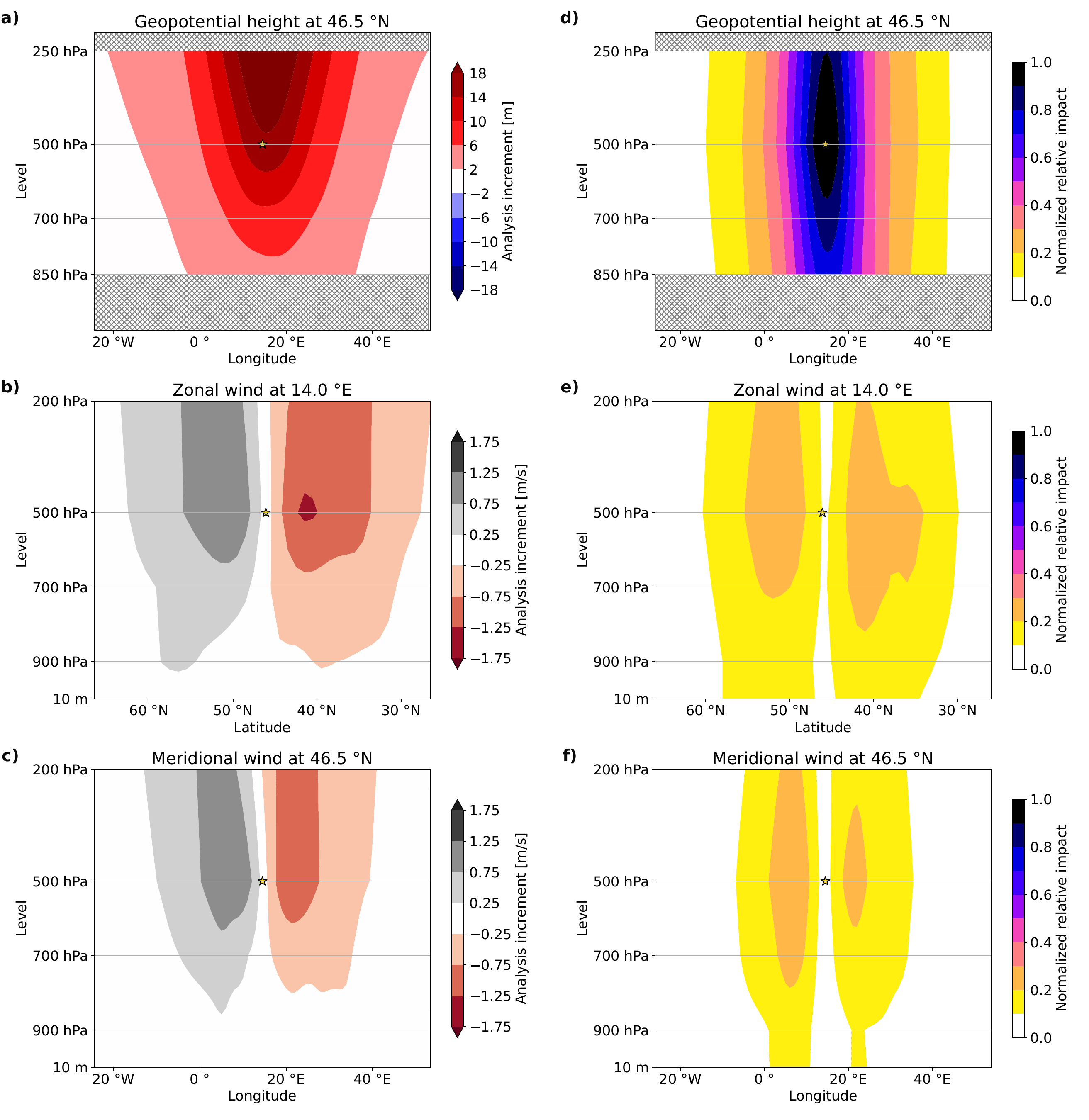}
    \caption{Vertical cross sections of analysis increments following an assimilation of Z500 observation. The cross section is done at the grid latitude or longitude nearest to the observation. (a) 2D longitude-pressure cross section of geopotential height increment at latitude $46.5\,^\circ\mathrm{N}$. 
    (b) 2D latitude-pressure cross section of zonal wind increment at longitude $14.0\,^\circ\mathrm{E}$. (c) 2D latitude-pressure cross section of meridional wind at latitude $46.5\,^\circ\mathrm{N}$. (d-f) As (a-c), but showing the normalized relative impact of the observation.
    Gaussian filtering with a standard deviation of 1\,$^\circ$ was applied in both horizontal directions to smoothen the contours. The observation location is denoted by a golden star.}
    \label{fig:Ljubljana_crossection}
\end{figure}

The geopotential height increment extends vertically (Fig.~\ref{fig:Ljubljana_crossection}a), with its magnitude increasing with elevation. This is expected, as geopotential height is a vertical integral of temperature, and the positive temperature increments (shown in Fig.~\ref{fig:Ljubljana}b,c) accumulate with height.
To measure the relative impact of the observation on the analysis, we define it as
$
    \mathcal{I}^\mathrm{V} =  \|\delta_a^\mathrm{V}\| / \sigma_a^\mathrm{V},
$
where $\mathrm{V}$ is the variable of interest. Figure~\ref{fig:Ljubljana_crossection}d-f shows the crossections of $\mathcal{I}^\mathrm{Z500}$, $\mathcal{I}^\mathrm{U500}$, and $\mathcal{I}^\mathrm{V500}$, normalized with $\mathcal{I}^\mathrm{Z500}$ value at the observation location. Despite Z500 analysis increment increasing with height, the relative impact of the observation peaks at the observation level, and diminishes with distance (Fig.~\ref{fig:Ljubljana_crossection}d). Similar patterns for increment magnitude and observation impact are observed for both horizontal wind components (Fig.~\ref{fig:Ljubljana_crossection}b,c,e,f).


\begin{figure}[h]
    \centering
    \includegraphics[width=\textwidth]{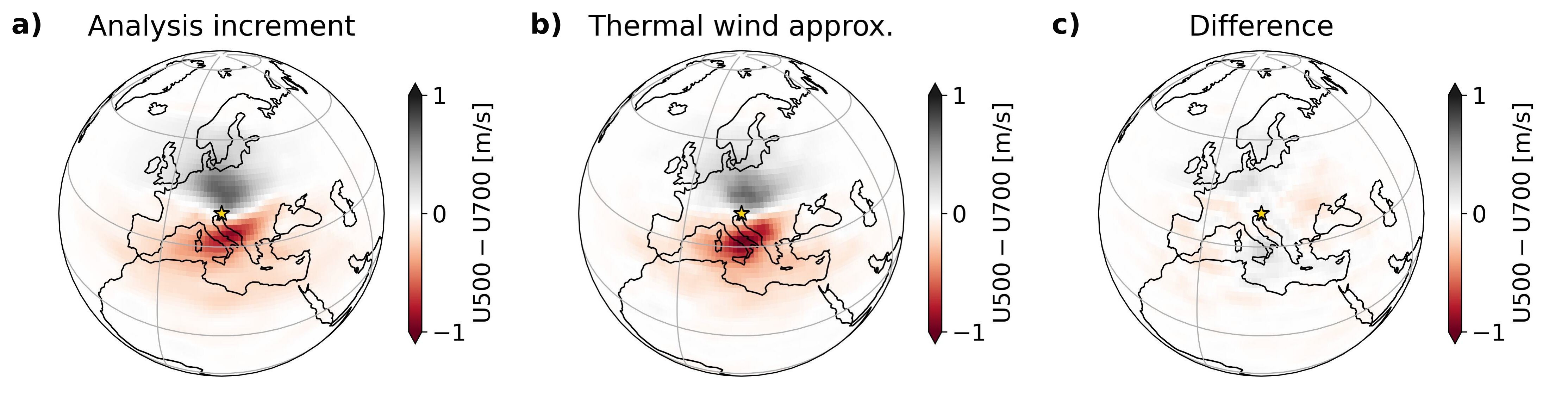}
    \caption{A comparison of (a) the difference in U500 and U700 analysis increments to (b) their difference derived from the thermal wind approximation in Eq.~\eqref{eq:thermal wind}. (c) The difference ((a)$-$(b)).}
    \label{fig:thermal_wind_balance}
\end{figure}

So far, we have shown that horizontal balance in the midlatitudes is well preserved following a single observation assimilation, whereas in the vertical direction, the impact of the observation
shows a maximum at the 
observed level and gradually decreases both upward and downward. Another key physical constraint in this region is the thermal wind balance~\citep{Holton2013AnMeteorology}, which links the vertical variation of horizontal wind to horizontal gradients in thickness (difference in geopotential height), assuming geostrophic balance in the horizontal and hydrostatic balance in the vertical.
Under these assumptions, the difference in analysis increments of zonal wind between 500\,hPa and 700\,hPa can be approximated by the thermal wind relation:
\begin{equation}
\label{eq:thermal wind}
    \delta^{\mathrm{U500}}_{a} - \delta^{\mathrm{U700}}_{a} \approx - \frac{g}{f} \frac{\partial (\delta^\mathrm{Z500}_a - \delta^\mathrm{Z700}_a)}{\partial y},
\end{equation}
where $g$ is acceleration due to gravity, $f$ is the Coriolis parameter, and $\partial/\partial y$ denotes the meridional derivative in units m$^{-1}$.
Figure~\ref{fig:thermal_wind_balance} confirms that this approximation holds well for the computed analysis increments, indicating that both the hydrostatic and the geostrophic balances are preserved after the assimilation.

\begin{figure}[h!]
    \centering
    \includegraphics[width=\linewidth]{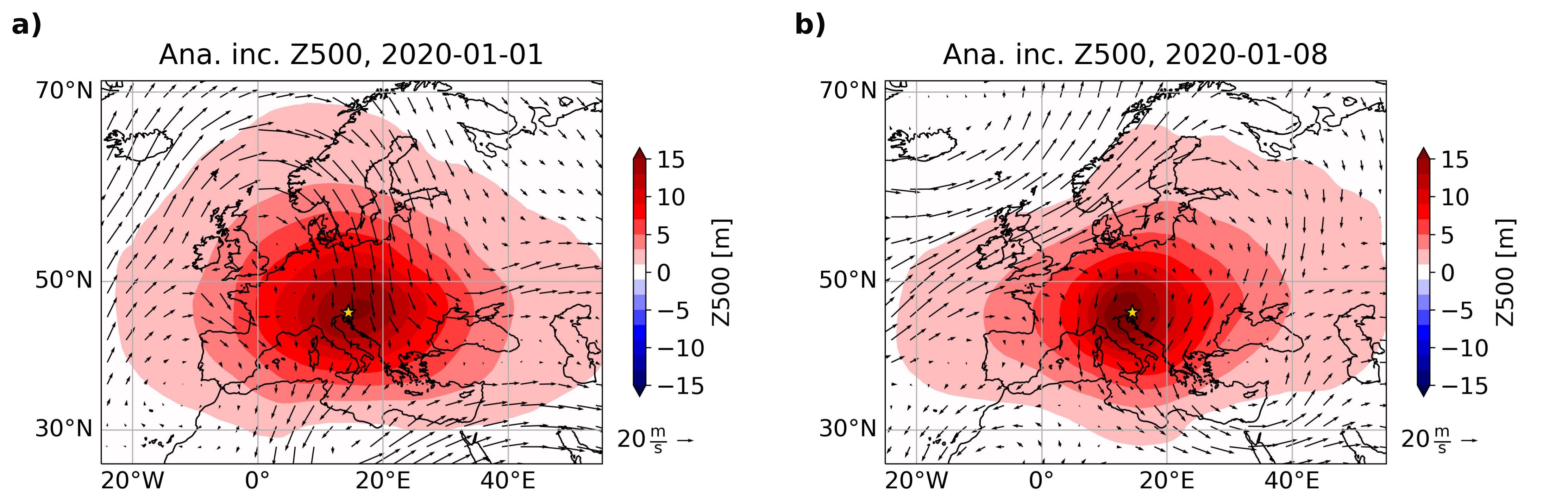}
    \caption{Z500 analysis increments following assimilation of Z500 observation above Ljubljana with 30\,m departure and 10\,m standard deviation on two different dates with different backgrounds: (a) January 1st, 2020, at 00 UTC, and (b) January 8th, 2020, at 00 UTC. The arrows denote the background 500\,hPa wind. The observation location is marked by a golden star.}
    \label{fig:different date}
\end{figure}

An important feature of the latent-space DA system is that the analysis increment depends on the background state due to the decoder's nonlinearity (MZ24), making the analysis increment effectively flow-dependent. This is demonstrated by repeating the same single-observation experiment on two different dates (Fig.~\ref{fig:different date}). For instance, the increment in Fig.~\ref{fig:different date}a is stretched meridionally, reflecting the more meridional background flow near the observation.

\begin{figure}[h]
    \centering
    \includegraphics[width=\textwidth]{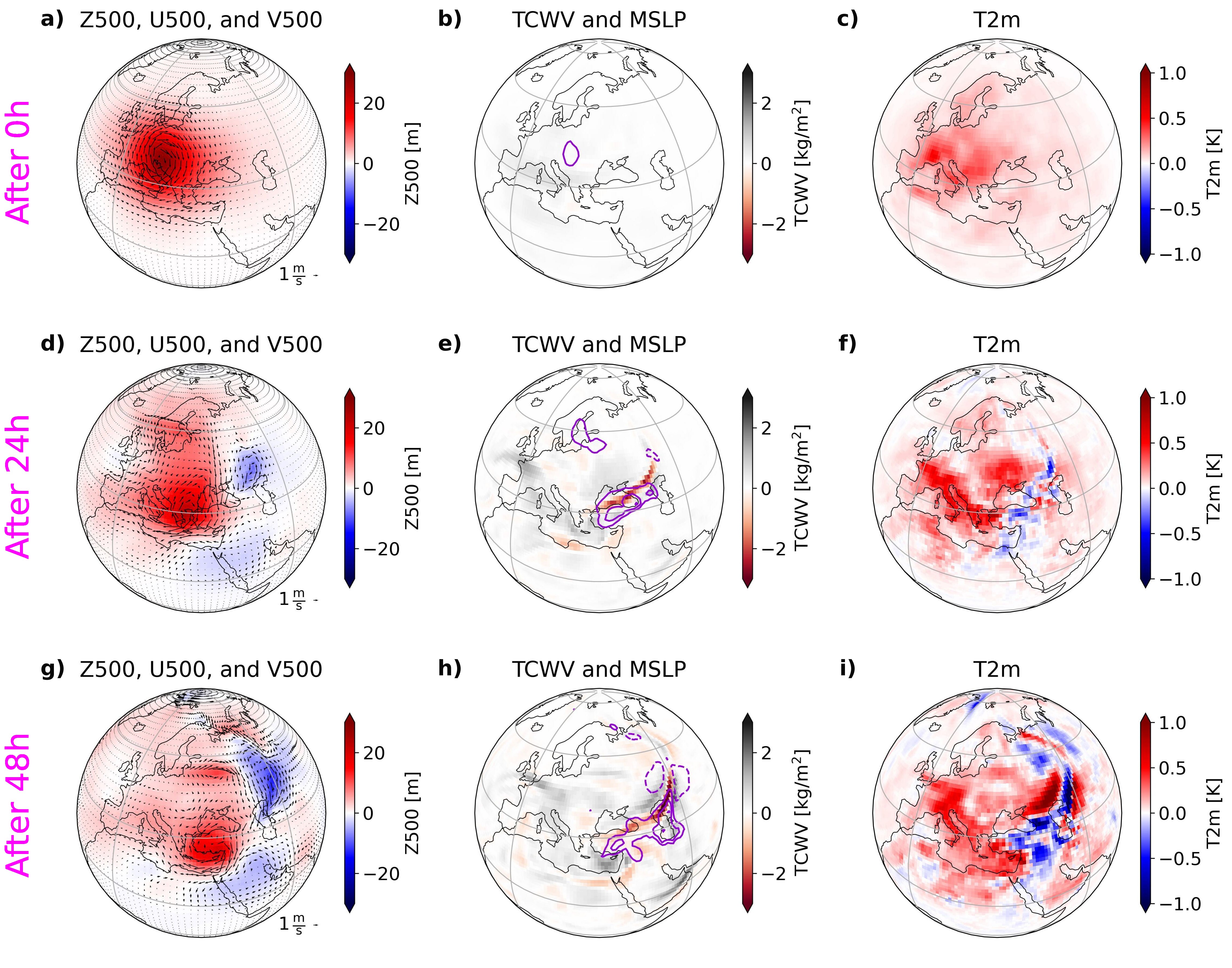}
    \caption{Difference between the two forecasts initialized from the analysis and the background, respectively, for a selected ensemble member, based on the experiment in Fig.~\ref{fig:Ljubljana}. Difference in the initial condition (the analysis increment) for (a) Z500 (colors) and 500\,hPa wind (arrows), (b) total-column water vapor (colors) and MSLP (purple contours), and (c) T2m. (d-f) As (a-c), but for the 24-hour forecast lead time. (g-i) As (d-f), but for the 48-hour forecast lead time. The solid/dashed contours in (b,e,h) indicate a positive/negative difference with 0.5\,hPa step, zero contour is omitted.}
    \label{fig:adjustment}
\end{figure}

No data assimilation algorithm yields a perfectly balanced analysis, so inconsistencies with the model's internal dynamics can persist. As a result, assimilated information may introduce biases, dissipate during integration, or produce unrealistic forecasts~\citep{Kalnay2002AtmosphericPredictability}. Figure~\ref{fig:adjustment} illustrates the model’s response to modified initial conditions from a single-observation experiment, presenting the difference between two 48-hour forecasts initialized from the analysis and the background for the same ensemble member. 

A positive increment in Z500 at the initial time propagates eastward (Fig.~\ref{fig:adjustment}a,d,f), gradually also decreasing in magnitude. The increased anticyclonic vorticity also leads to increased advection of negative vorticity downstream, leading to enhanced cyclonic vorticity at 24-hour lead time east of the initial perturbation.
To the southeast of the propagating signal, MSLP increases as expected, accompanied by a reduction in TCWV (Fig.~\ref{fig:adjustment}e,h). This moisture reduction slightly lags a concurrent drop in T2m in the same region (Fig.~\ref{fig:adjustment}f,i), collectively resembling the atmospheric response to a passing cold front. Overall, the primary information introduced by the analysis increment is retained throughout the 48-hour forecast integration.

\subsection{Background error-covariance model in the tropics -- a response to humidity saturation}
\label{sec:humidity saturation}

The physical balances in the tropics vastly differ from those in the midlatitudes due to a much smaller Coriolis force and tend to be too complex for analytical description. The primary energy source for atmospheric motions in the tropics is the condensation heating in convective cloud systems which drives the large-scale tropical circulation~\citep{Holton2013AnMeteorology}. The diabatic heating is balanced by the adiabatic cooling of ascending flow (updrafts) inside. This leads to a drop in the surface pressure below the perturbation, establishing a convergent horizontal wind pattern in the lower troposphere that can further fuel the development of a convective system. At the top of the convective system, the statically-stable tropopause acts as a lid for strong vertical motions, leading to the positive pressure perturbation and divergent horizontal outflow~\citep{Gill1980SomeCirculation,Davey1987ExperimentsModel}. 

\begin{figure}[h]
    \centering
    \includegraphics[width=\textwidth]{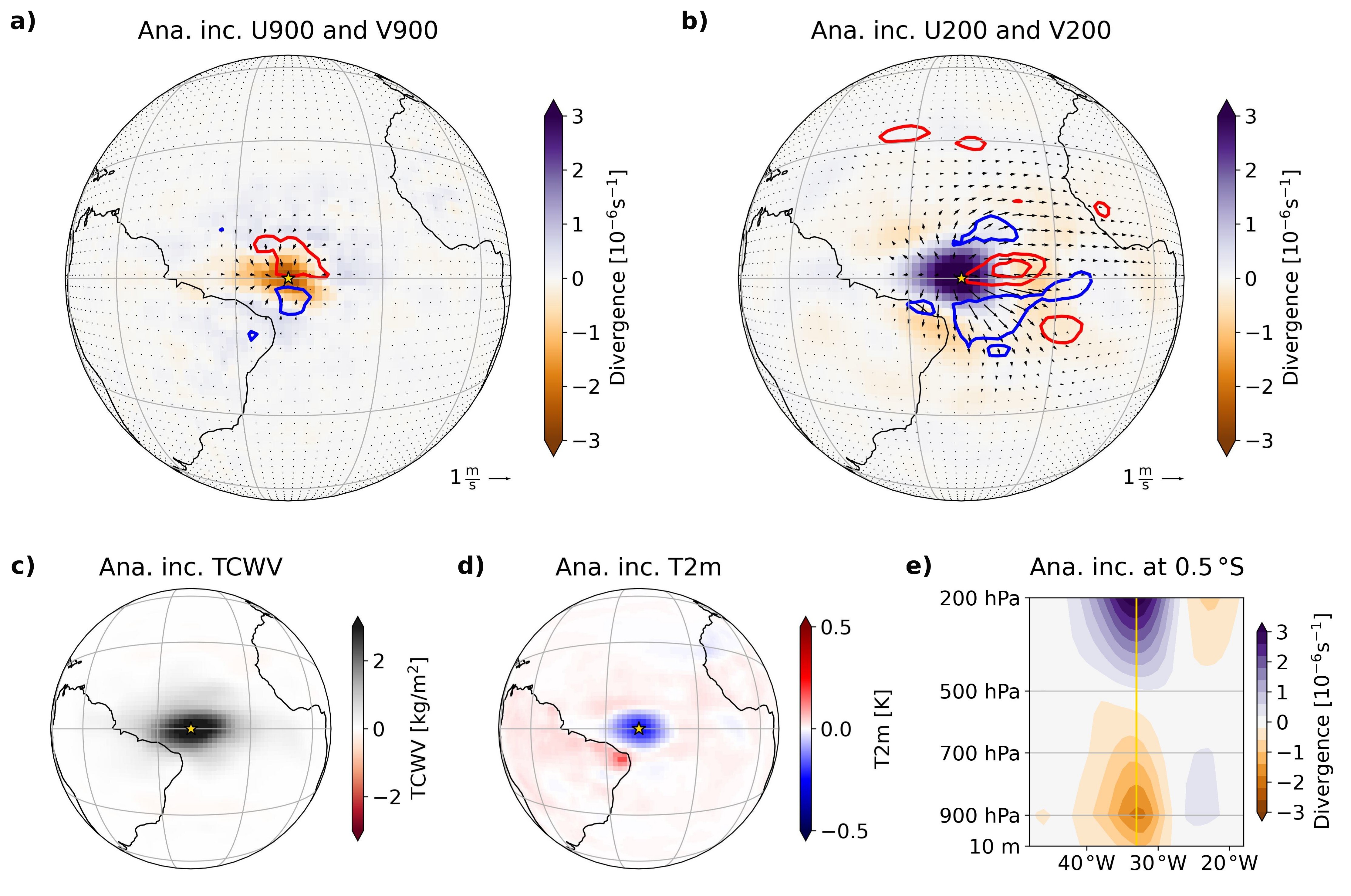}
    \caption{Analysis increments following an assimilation of TCWV observation above Central Atlantic with 10\,kg/m$^2$ departure and 3\,kg/m$^2$ observation-error standard deviation. (a) 900\,hPa horizontal wind increment (arrows), its divergence (colors) and vorticity (contours); (b) as (a) but at 200\,hPa; (c) TCWV increment; (d) T2m increment; (e) 2D longitude-pressure cross section of divergence of the wind increments at the 
    observation location ($0.5\,^\circ\mathrm{S}$). The vertical line corresponds to the longitude of the observed TCWV. Red/blue vorticity contours in (a,b) correspond to positive/negative vorticity with a step of $5\times10^{-7}$\,s$^{-1}$, zero-contour is omitted. Horizontal smoothing of the contours in panel (e) was applied as in Fig.~\ref{fig:Ljubljana_crossection}.}
    \label{fig:TCWV}
\end{figure}

Although our applied neural networks do not explicitly resolve clouds, condensation, latent heat release, or precipitation, they relate total column water vapour (TCWV) to other atmospheric variables. We tested whether an increase in TCWV in a tropical area with high background TCWV leads to similar effects as described in \citet{Davey1987ExperimentsModel} and obtained in simplified numerical studies of tropical response to latent heating~\citep{Zaplotnik2018AnProcesses}. Figure~\ref{fig:TCWV} shows the analysis increments following an assimilation of TCWV observation in the Central Atlantic Ocean (equator, $33.0\,^\circ\mathrm{W}$) with departure $d^\mathrm{TCWV} = 10$\,kg/m$^2$ and error standard deviation $\sigma^\mathrm{TCWV}_o = 3$\,kg/m$^2$. The horizontal wind increments at 900\,hPa (Fig.~\ref{fig:TCWV}a) and 200\,hPa (Fig.~\ref{fig:TCWV}b) display convergence in the lower troposphere and divergence in the upper troposphere. At 900\,hPa, we also observe a cyclonic vorticity pattern north/south of the perturbation. At 200\,hPa, the wind increments are substantially stronger, and the latent-space DA system even generates a Kelvin wave east of the observation. The analysis increments that span the whole tropospheric column are a consequence of tropospheric-wide background-error covariances that are typical of precipitating regions where condensational latent heating occurs~\citet{Li2023ComparisonSystem}.

The analysis increment of TCWV (Fig.~\ref{fig:TCWV}c) has an elliptic shape with a greater extent in the zonal than in the meridional direction, reflecting the dominance of the zonal flow in the tropical atmosphere. A negative increment in T2m is also present (Fig.~\ref{fig:TCWV}d). This may indicate latent cooling due to evaporation of precipitation in the lowest layers -- once again implicitly captured by the autoencoder.

Does our DA and the associated latent-space $\mathbf{B}_z$-matrix capture the response to latent-heat-induced perturbations in the tropics? To provide further evidence, we analyze the vertical profile of horizontal divergence in the analysis increment. Figure~\ref{fig:TCWV}e shows that, at the location of the observed TCWV, the convergent wind increment is present in the lower troposphere, peaking around 900\,hPa. Near 500\,hPa, the flow transitions to divergent, which peaks at 200\,hPa, with a magnitude twice as large as the maximum convergence. This vertical structure closely resembles observational findings from \citet{Williams1973StatisticalPacific}, suggesting that our system realistically represents the dynamical response to tropical latent heating. This highlights a remarkable capability of the latent-space background-error covariance model -- one not yet demonstrated by any other variational DA system.

\begin{figure}[h!]
    \centering
    \includegraphics[width=.9\textwidth, clip, trim={0cm 1cm 0cm 0cm}]{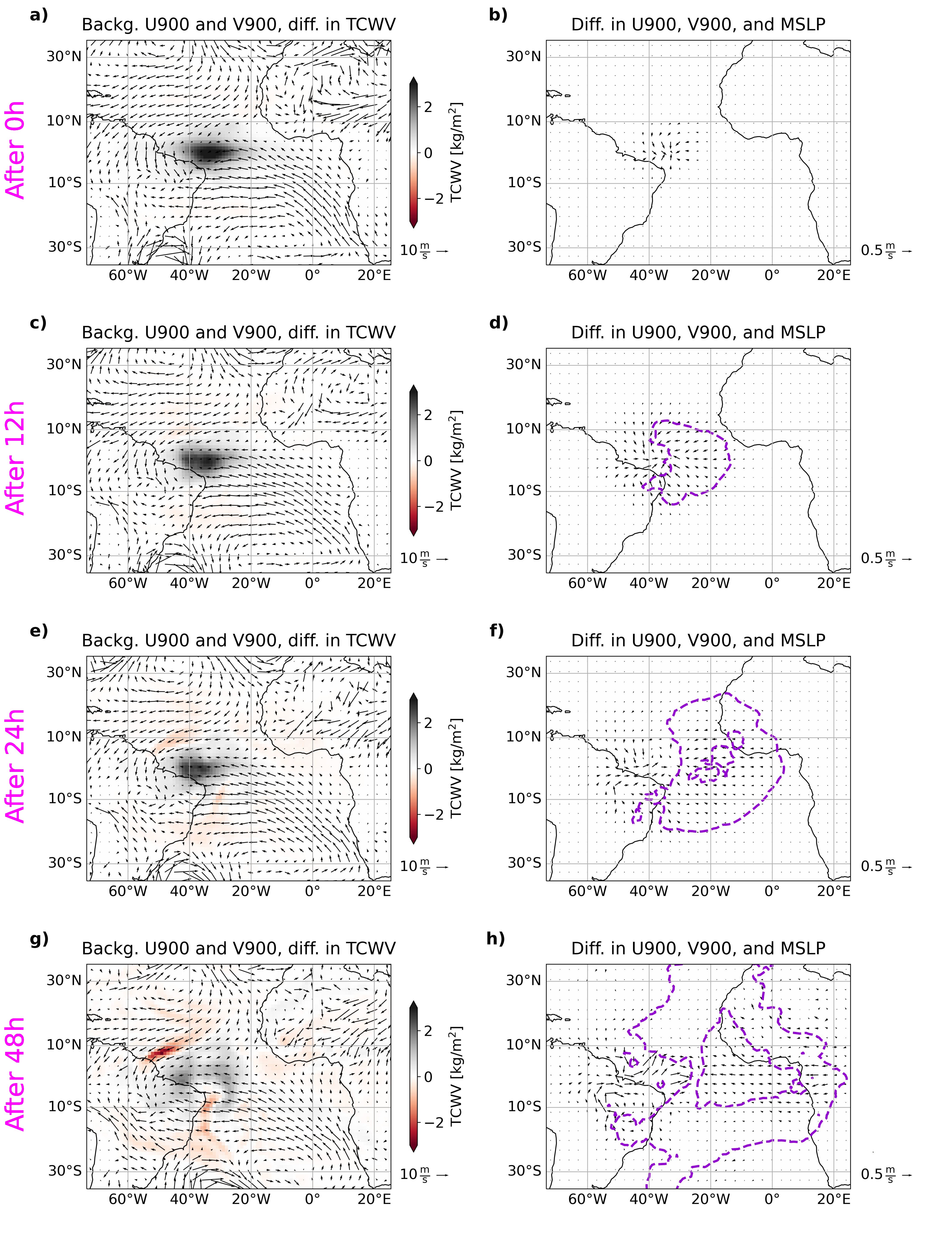}
    \caption{
    Difference between the two forecasts initialized from the analysis and the background from the single-observation experiment in Fig.~\ref{fig:TCWV}. (a) Difference in the initial condition for TCWV (colors) on top of the 900\,hPa background wind (arrows). (b) Difference in the initial condition for the 900\,hPa wind (arrows; there is an order of magnitude difference in wind magnitude scaling compared to panel (a)). (c-d) As (a-b), but for the 12-hour forecast lead time. (e-f) As (a-b), but for the 24-hour forecast lead time. (g-h) As (a-b), but for the 48-hour forecast lead time.. The dashed contours in (d,f,h) indicate a negative MSLP difference with 0.1\,hPa step, zero contour is omitted.}
    \label{fig:adjustment tropics}
\end{figure}

Similarly to Fig.~\ref{fig:adjustment}, Figure~\ref{fig:adjustment tropics} illustrates the model’s response to the initial perturbation introduced in the tropical single-observation experiment.  The TCWV increment is advected westwards with the background 900\,hPa wind (note that the 900\,hPa wind increment in Fig.~\ref{fig:adjustment tropics}b,d,f,h is negligible compared to the background shown in Fig.~\ref{fig:adjustment tropics}a,c,e,g). Since specific humidity decreases exponentially with height~\citep{Holloway2009MoistureConvection}, the TCWV advection is primarily governed by the lower-tropospheric winds. This behavior reaffirms the physical realism of the learned  neural-network dynamics. The positive TCWV increment disperses before reaching the coast of Brazil. Meanwhile, the 900\,hPa wind and MSLP increments (Fig.~\ref{fig:adjustment tropics}b,d,f) reveal the development of an eastward travelling Kelvin wave, which is fully developed by 48 hours (Fig.~\ref{fig:adjustment tropics}h). 

\goodbreak

\section{Estimating latent space background-error covariance model from operational EDA}
\label{sec:true EDA}

So far, we have shown that latent-space DA, using a climatological $\mathbf{B}_z$-matrix, produces physically reasonable analysis increments both in the tropics and midlatitudes. In this section, we extend our approach by incorporating the ensemble of data assimilation (EDA, \citep{Isaksen2010EnsembleECMWF}) to capture fully flow-dependent background-error variances, commonly referred to as "the errors of the day"~\citep{Bonavita2012On4D-Var}.


We used an ensemble of backgrounds, valid on April 14th, 2024, at 00\,UTC, from ECMWF's IFS cycle 49r1~\citep{ifs49r1DA}, to derive flow-dependent background-error covariances in the latent space. The ensemble-derived background-error covariance matrix was computed as
\begin{equation}
    \label{eq:Bz EDA}
    \mathbf{B}^\mathrm{EDA}_z =
    \frac{1}{N-1} \sum_{i=1}^N
    \Big(\mathbf{z}_{b,i}^\mathrm{IFS} - \left\langle\mathbf{z}_b^\mathrm{IFS}\right\rangle\Big) \Big(\mathbf{z}_{b,i}^\mathrm{IFS} - \left\langle\mathbf{z}_b^\mathrm{IFS}\right\rangle\Big)^\top,
\end{equation}
where $N=50$ is the total number of ensemble members, $\mathbf{z}_{b,i}^\mathrm{IFS} = E\circ S\left(\mathbf{x}_{b,i}^\mathrm{IFS}\right)$ is an encoded ensemble member with index $i$, and $\left\langle\mathbf{z}_b^\mathrm{IFS}\right\rangle = \frac{1}{N}\sum_{i=1}^N E\circ S\left(\mathbf{x}_{b,i}^\mathrm{IFS}\right)$ is the ensemble mean of the encoded vectors. Figure~S5 shows that the latent-space variances (diagonal elements) typically exceed the covariances (off-diagonal elements) by an order of magnitude. However, this contrast is less pronounced than in the climatological $\mathbf{B}_z$ (Fig.~S3b), which makes neglecting off-diagonal elements in 3D-Var minimization less justifiable. However, the off-diagonals span a similar range in both matrices, which may suggest they primarily reflect sampling noise. 

Climatology-based background-error variances tend to overestimate forecast uncertainty~\citep{Bannister2008a}, whereas ensemble-derived variances often underestimate it. 
The vast difference in variances is evident in Fig.~S6, which shows that the variances from the operational ensemble are approximately an order of magnitude smaller than those in $\mathbf{B}_z^{clim}$ (compare to Fig.~S4, rows C and H). Moreover, neural networks tend to smooth meteorological fields~\citep{Bonavita2024OnModels}, reducing their ability to reconstruct the fine-scale features after encoding and decoding. For example, Fig.~S6 displays how the standard deviation of the 50 ensemble members changes after autoencoder processing, with noticeable loss of fine-scale details due to the AE’s limited resolution. While ensemble variance is reasonably preserved for the geopotential height, it is poorly reconstructed for fields like winds and specific humidity, which contain greater small-scale variance. 

\begin{figure}[h!]
    \centering
    \includegraphics[width=\linewidth]{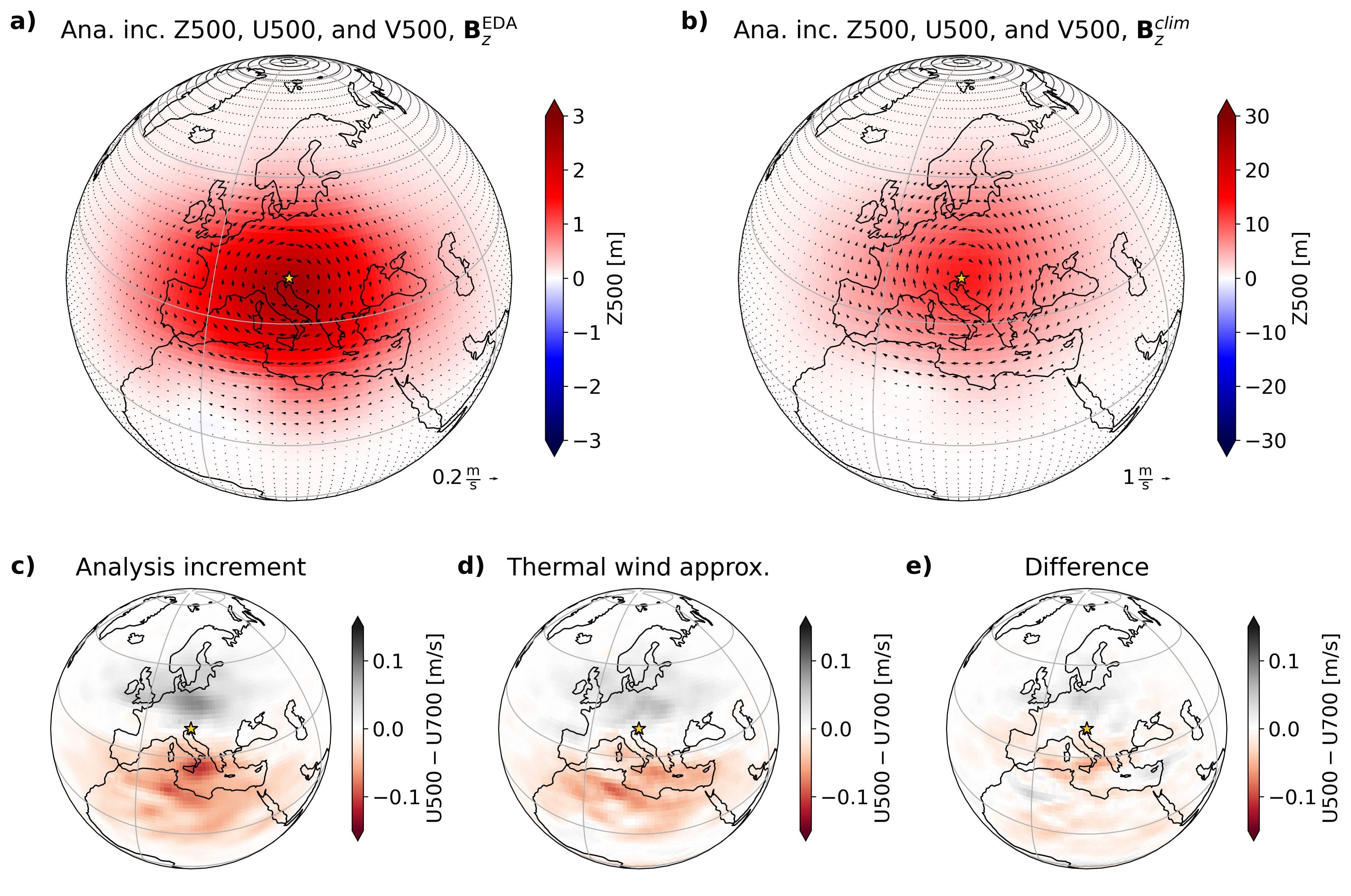}
    \caption{(a,b) As Fig.~\ref{fig:Ljubljana}a but for an experiment using an operational EDA ensemble of backgrounds and (a) its corresponding $\mathbf{B}_z$-matrix and (b) climatological $\mathbf{B}_z$-matrix from Eq.~\eqref{eq:Bz clim}. Note the different color and arrow scales in panel (a) compared to Fig.~\ref{fig:Ljubljana}a and Fig.~\ref{fig:true EDA}b. (c-e) As Fig.~\ref{fig:thermal_wind_balance}a-c, but computed for the experiment from panel (a).}
    \label{fig:true EDA}
\end{figure}

Using an ensemble of encoded backgrounds from EDA and EDA-derived background-error covariance matrix (Eq.~\ref{eq:Bz EDA}), we repeated a single-observation experiment with $d^\mathrm{Z500} = 30\,\mathrm{m}/\mathrm{s}$ and $\sigma^\mathrm{Z500}_o = 10\,\mathrm{m}/\mathrm{s}$ over Ljubljana. Figure~\ref{fig:true EDA}a,c-e reveals that the resulting analysis increment preserved the geostrophic and thermal wind balance. 
To isolate the effect of the $\mathbf{B}_z$-matrix from the impact of using operational ensemble of backgrounds, we repeated the experiment once more using the same ensemble of backgrounds, but applied $\mathbf{B}^{clim}_z$ for 3D-Var cost function computation (Fig.~\ref{fig:true EDA}b). Qualitatively, the analysis increments exhibited only minor differences. However, the magnitude of the analysis increment was significantly larger using $\mathbf{B}^{clim}_z$ (approximately five times larger at the observation location), while the background-error standard deviation was larger too (fourfold increase; not shown), both of which are expected due to an order of magnitude difference in the latent-space variances between the respective $\mathbf{B}_z$ matrices.

\begin{figure}[h!]
    \centering
    \includegraphics[width=\linewidth]{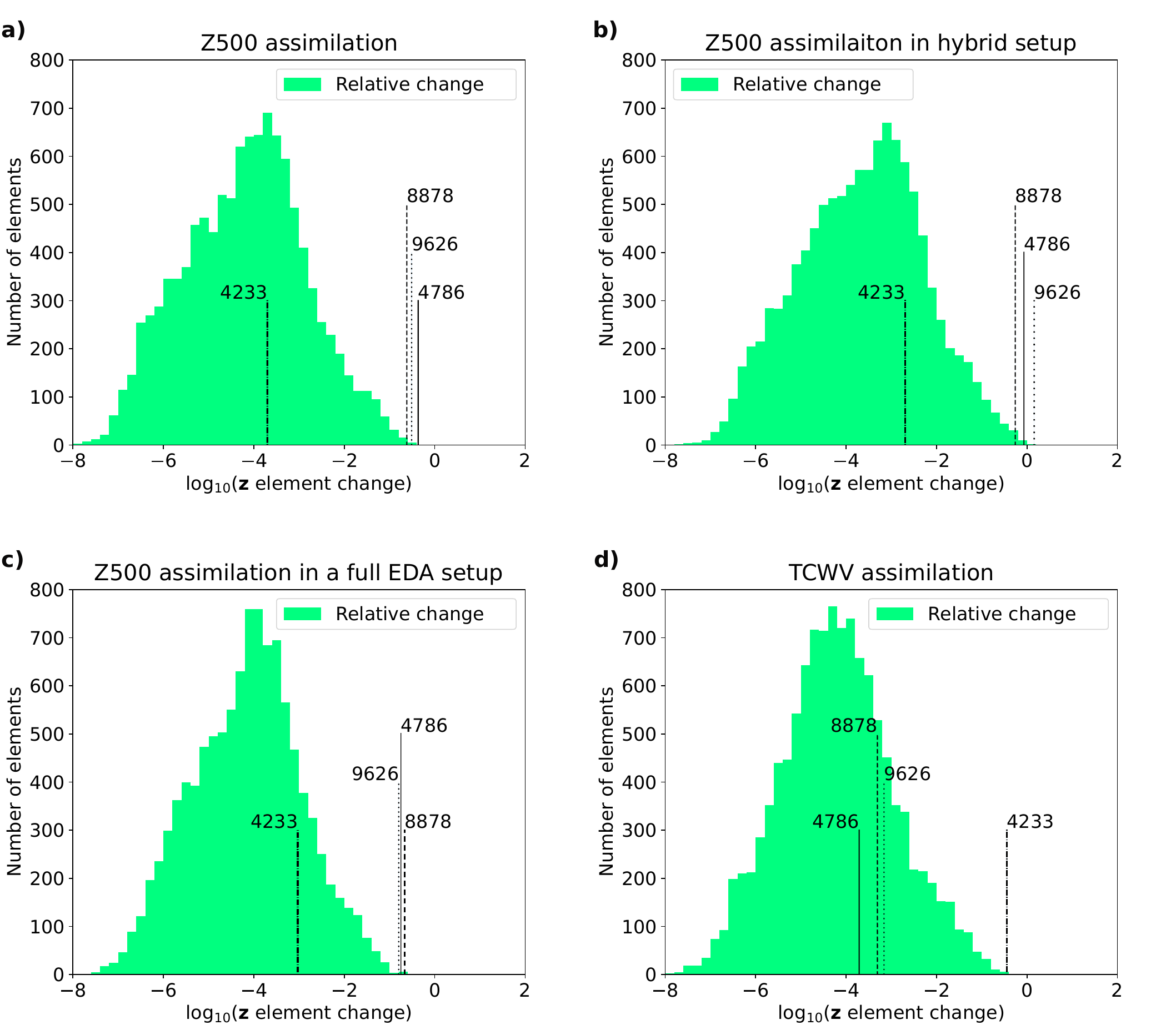}
    \caption{Histograms of the relative changes of latent vector elements in different single-observation experiments. 
    The indices of the most altered element in each of the experiments are highlighted in all four plots. (a) The histogram for the experiment with observing Z500 above Ljubljana using the original setup (Sec.~\ref{sec:geostrophic adjustment}). (b) The histogram for the experiment with observing Z500 above Ljubljana using the operational ensemble of backgrounds and climatological~$\mathbf{B}_z$ in 3D-Var cost function minimization. (c) The histogram for the experiment with observing Z500 above Ljubljana using IFS ensemble forecast as the background and its proper~$\mathbf{B}_z$ in 3D-Var cost function. (d) The histogram for the experiment with observing TCWV above Central Atlantic using the original setup (Sec.~\ref{sec:humidity saturation}). The bin width is 0.2 on a logarithmic scale.}
    \label{fig:changes in latent space}
\end{figure}

Beyond the differences in the physical space, the analysis increments also differ in the latent space. Figure~\ref{fig:changes in latent space} presents histograms of the relative changes in latent vector elements (i.e., ratio of the analysis increment and standard deviation) following 3D-Var cost function minimization across four different experiments. These histograms reveal that, in each experiment, only a small fraction of 12100 elements undergo substantial modification. In the three experiments involving single Z500 observation, the same subset of latent elements is consistently among the most modified, and these changes predominantly affect the region where the analysis increment peaks (Fig.~S7). 
In contrast, these elements are negligibly modified when observing TCWV in the Central Atlantic, where the most adjusted latent element mainly alters the region in the vicinity of that observation (Fig.~\ref{fig:changes in latent space}d, Fig.~S7, row J).

\begin{figure}[h!]
    \centering
    \includegraphics[width=\linewidth]{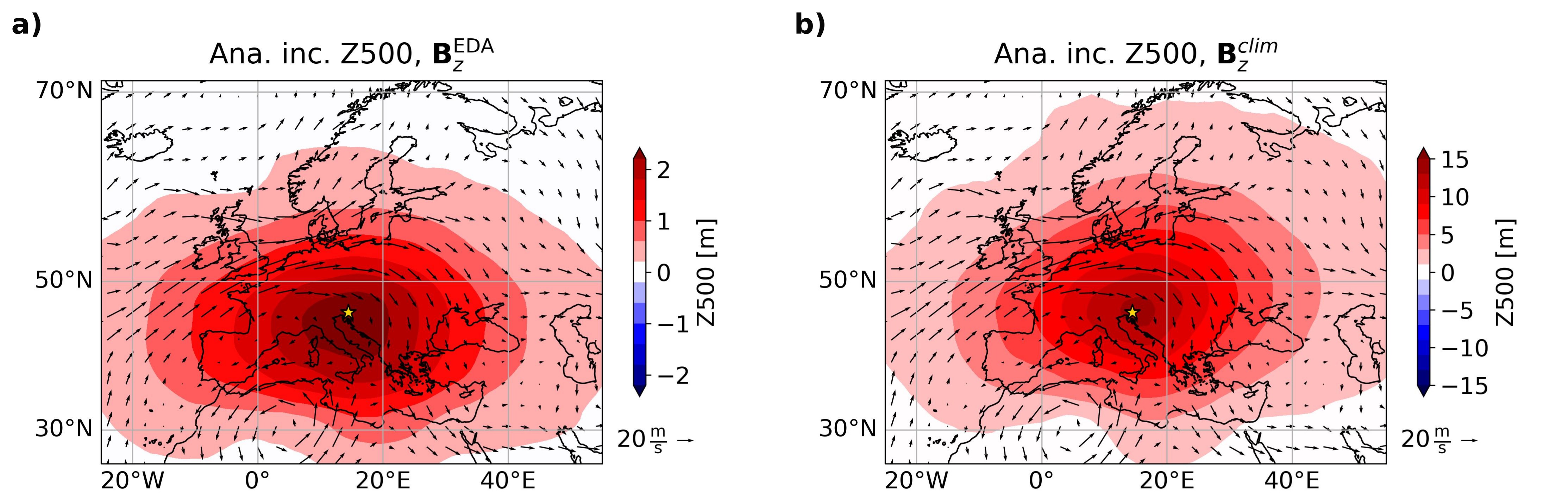}
    \caption{As Fig.~\ref{fig:different date}, but using the encoded IFS ensemble as the background and (a) $\mathbf{B}_z^\mathrm{EDA}$ or (b) $\mathbf{B}_z^{clim}$ in 3D-Var cost function computation.}
    \label{fig:flow dependence}
\end{figure}

The most important benefit of using operational forecast ensembles in DA is their ability to provide flow-dependent background-error variances. \citet{Fan2025PhysicallySpace} demonstrated that even when using a static, climatology-based $\mathbf{B}_z$-matrix (in their case computed via the NMC method), latent DA can still produce covariances in the physical space that reflect the background state, and we reaffirmed that in Fig.~\ref{fig:different date}. Losing this capability when transitioning from $\mathbf{B}_z^{clim}$ to an ensemble-derived $\mathbf{B}_z^\mathrm{EDA}$ would represent a major drawback and would strongly undermine the motivation for making such a change.
Similarly to Figure~\ref{fig:different date}, Figure~\ref{fig:flow dependence} illustrates two examples of the impact of a single Z500 observation above Ljubljana, but using the same ensemble of background states in both cases and different $\mathbf{B}_z$-matrices. While it is difficult to determine which spread of the information from the observation more faithfully follows the background flow, both retain some flow-dependent structure. This indicates that switching from $\mathbf{B}_z^{clim}$ to $\mathbf{B}_z^\mathrm{EDA}$ does not degrade the method’s ability to represent flow-dependent features. However, given the current performance of the AE, it also does not lead to substantial improvements.

\section{Discussion, conclusion, and outlook}
\label{sec:discussion}

The motivation for this study was to advance the representation of background-error covariances for variational data assimilation (DA) to better capture both tropical and extratropical balances. Although several analytical approaches have been proposed to address this challenge~\citep{Zagar2004VariationalConstraint, Koernich2008Combining}, none have been implemented in operational systems. Therefore, they still lack multivariate representation of tropical background-error covariances, including the balance structures. To address this gap, we explored an alternative approach by computing covariances in the reduced-dimension latent space, learned by a convolutional autoencoder, and reformulating the 3D-Var cost function to be minimized within that space. This method has already shown promising results in a univariate context (MZ24), and has been demonstrated to produce analysis increments in the midlatitudes that exhibit both geostrophic horizontal structure and flow-dependent characteristics~\citep{Fan2025PhysicallySpace}. In the present study, we extended the approach in MZ24 to multivariate, multilevel representation of the atmosphere, and tested the physical plausibility of the resulting analysis increments through single-observation data assimilation experiments.

Using a climatology-based background-error covariance matrix computed as a difference between 24-hour neural network forecast and the corresponding ground truth, we showed that an observation of 500\,hPa geopotential height (Z500) produces an analysis increment with a geostrophically balanced pattern in both 500\,hPa geopotential height and winds (Fig.~\ref{fig:Ljubljana}a). The observational information is also realistically propagated vertically and across variables (Figs.~\ref{fig:Ljubljana}b,c,~\ref{fig:Ljubljana_crossection}). Moreover, the analysis increment preserves the thermal wind balance (Fig.~\ref{fig:thermal_wind_balance}), indicating that both geostrophic and hydrostatic balance are preserved during the assimilation process.
In the tropics, an assimilation of a single total column water vapor (TCWV) observation with positive departure in an atmosphere near saturation led to an analysis increment consistent with the theoretical dynamical response to heat-induced perturbations (Fig.~\ref{fig:TCWV}), resulting in convergent winds in the lower troposphere and divergent winds aloft, a fully multivariate response that is currently not captured by operational variational DA systems. This happened despite the humidity, convection, and precipitation not being explicitly represented in the applied neural networks.
Forecasts initialized from the resulting analyses produced physically plausible weather evolution: in the midlatitudes, a developing cold front (Fig.~\ref{fig:adjustment}); and in the tropics, a propagating Kelvin wave and a TCWV structure advected by lower-tropospheric winds (Fig.~\ref{fig:adjustment tropics}).

A key difference between the analysis increments in this study and those presented in MZ24 lies in their horizontal extent -- here, the increments are significantly more localized, particularly in the tropics. We found that perturbing individual elements of the latent vector leads to localized changes in the decoded atmospheric fields (Fig.~S7). During assimilation, the algorithm primarily adjusted those latent vector elements that describe the features near the observation location (Figs.~\ref{fig:changes in latent space},~S7). This represents a substantial improvement in the applicability -- unlike MZ24, where perturbing individual latent vector elements caused global changes in the decoded fields. This improvement is likely due to the use of a higher-quality autoencoder in the present study, featuring a latent space two orders of magnitude larger than that of MZ24. Incorporating neural networks with even greater representational capacity and larger latent spaces could further enhance localization and provide a more diverse set of learned weather patterns within the latent space.

The presented method may also be promising for representing background-error cross-covariances between atmosphere, ocean, and surface, potentially enabling a unified framework for assimilating ocean, land, and even hydrological observations -- traditionally handled separately~\citep{Park2009DAAtmosphereOceansHydrology}. While \citet{Zheng2024GeneratingFramework} applied a similar method to assimilate sea surface temperature over the Pacific, we explored how surface temperature (ST) observations  over land and ocean affect the local vertical temperature profile (not shown). However, we did not observe substantial differences, likely due to limitations in the reconstruction quality of our NNs and coarse vertical resolution of temperature fields. Overall, the $\mathbf{B}_z$-matrix realistically captures the contrast in ST variance between land and sea (Fig.~S4, panel C6), and the assimilation procedure produces distinct responses in ST and 2-meter temperature (T2m). For example, assimilating Z500 over Ljubljana results in a T2m impact roughly twice that of ST (Fig.~S4, panels E6-7). Similarly, a positive TCWV departure over the central Atlantic generates a local negative increment appears in T2m (Fig.~\ref{fig:TCWV}d), while the ST increment shows no clear pattern.

This study also addressed the transition from climatology-based to ensemble-based background error covariance matrices. Despite a substantial variance loss when encoding the ECMWF's operational EDA  (Fig.~S6) due to limitations in the autoencoder's representational capacity, the analysis increment from assimilating Z500 in the midlatitudes still respected both horizontal and vertical physical balances. Crucially, the flow-dependent characteristics were preserved (Fig.~\ref{fig:flow dependence}). In the long term, using ensemble-derived background error covariances could better capture
flow-dependent features and estimate covariance magnitudes more accurately than climatology-based matrices. 

As prior latent-space DA studies~\citep[MZ24,][]{Zheng2024GeneratingFramework, Fan2025PhysicallySpace}, we used only the diagonal elements of the background-error covariance matrix in the 3D-Var cost function. Although, we did not explicitly address the role of off-diagonal terms, the resulting increments consistently respected physical balances across the $\mathbf{B}_z$ matrices tested. While higher-capacity neural networks could improve performance, they would require larger latent spaces -- making the quasi-diagonality assumption for $\mathbf{B}_z$ effectively unavoidable. Potential challenges from latent-space non-Gaussianity could be mitigated using techniques like normalizing flows~\citep{Boehm2022}.

Overall, this method has demonstrated the ability to generate physically consistent, flow-dependent analysis increments in both tropical and extratropical atmosphere, all while maintaining relatively low computational cost. Given its promising performance with ensemble-derived background-error covariances and the physical realism of forecasts initialized from the resulting analyses, we believe this approach could be further extended towards a full ensemble-based latent-space 4D-Var system, comparable to those used in operational weather forecasting centers.

\section*{Acknowledgements}
This research was funded by the Slovenian Research and Innovation Agency (ARIS) Programme
P1-0188, Grant MR\,56969, and Grant 0510--1554-58115. This research was also supported by the University of Ljubljana Grant SN-ZRD/22-27/0510. Žiga Zaplotnik acknowledges the funding by the European Union under the Destination
Earth initiative. 
The authors are grateful to Nedjeljka Žagar (Universität Hamburg) and Massimo Bonavita (ECMWF) for fruitful discussions on the topic.

\bibliographystyle{abbrvnat}
\bibliography{references}

\newpage

\appendix
\section{Appendix A: Neural network design}
\label{app:nn}
\renewcommand{\thefigure}{A\arabic{figure}}
\setcounter{figure}{0}
\subsection{Autoencoder}
\begin{figure}[h!]
    \centering
    \includegraphics[width=\textwidth, clip, trim={2cm 0.2cm 0cm 0cm}]{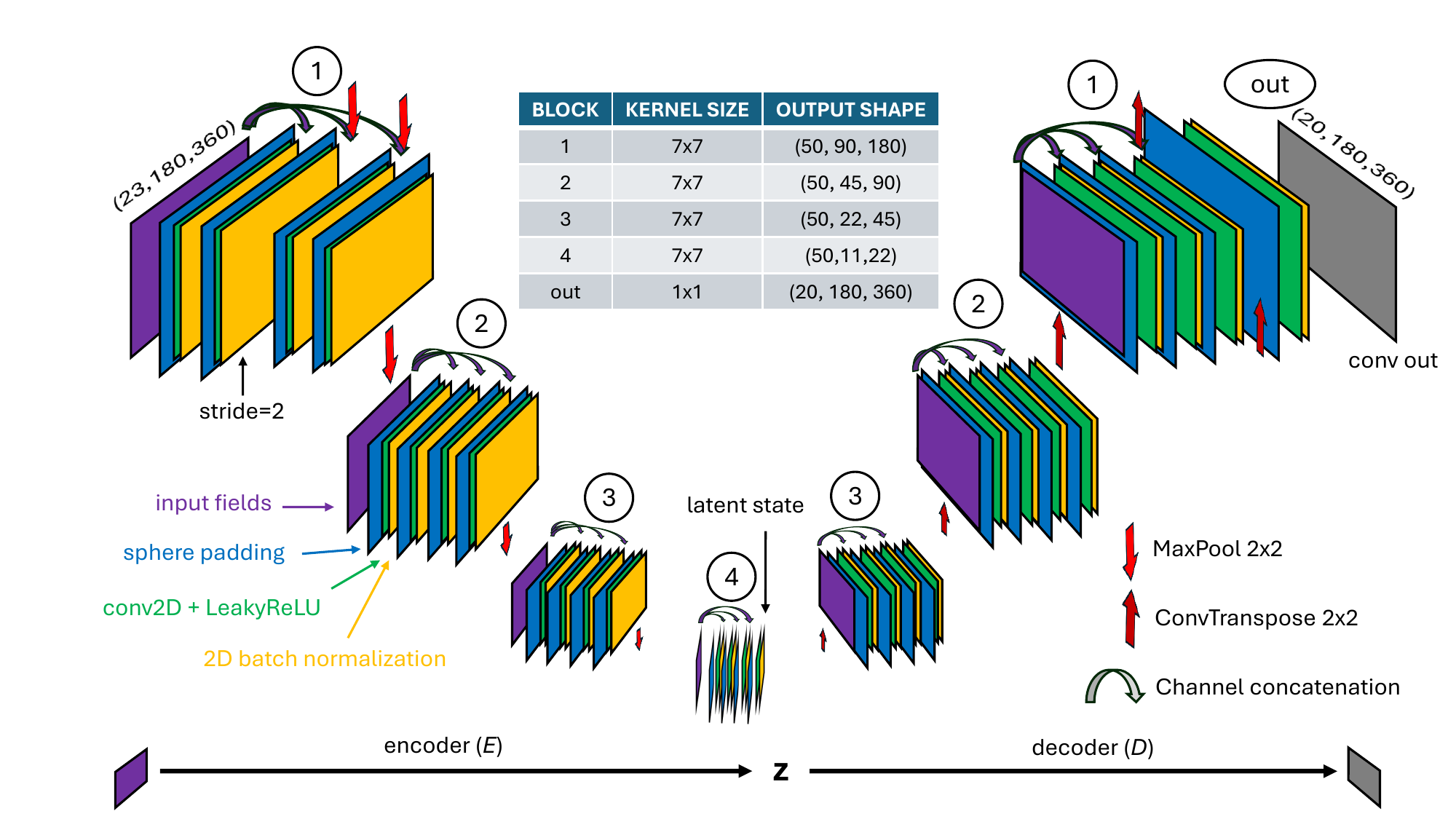}
    \caption{Autoencoder structure. The kernel sizes in the table correspond to the convolutional blocks in the downscaling/upscaling part of the network (left/right part of the scheme).}
    \label{fig:AE}
\end{figure}

The structure of the autoencoder (AE), illustrated in Fig.~\ref{fig:AE}, consists of the encoder and the decoder. The encoder $E$ comprises four convolutional blocks, each containing four subblocks with a consistent design. In each block, the input field is first padded using spherical padding, which is periodic in the zonal direction and polar in the meridional direction~\citep{Perkan2025UsingModels}. The padded field is then passed through a two-dimensional (2D) convolutional layer with a $7\times7$ kernel and 50 output channels, followed by a leaky rectified linear unit (Leaky ReLU) activation and 2D batch normalization. The stride is set to 1 in all convolutional layers, except in the second subblock of the first block, where it is set to 2. Before entering each subsequent subblock, the output is concatenated with the original input to the first subblock. After each block, the spatial dimensions are reduced using $2\times2$ max pooling.

The encoder's output has the shape $50\times11\times22$, defining the latent state. Its flattened form is referred to as the latent vector $\mathbf{z}$, consisting of 12100 elements. This vector is modified during the data assimilation procedure (Sec.~\ref{sec:3D-Var}). The latent state is then passed into the decoder $D$, which consists of three convolutional blocks that mirror the encoder’s structure, except that the final subblock in each block includes a transposed 2D convolutional layer that doubles the spatial dimensions in both latitude and longitude. The output of the final decoder block is passed through a pointwise ($1\times1$) convolutional layer, yielding a reconstructed atmospheric field of shape $20\times180\times360$. A qualitative comparison between an ERA5 reanalysis field and its corresponding autoencoded version is shown in Fig.~\ref{fig:AE quality}.

\subsection{Forecasting model}
\begin{figure}[h!]
    \centering
    \includegraphics[width=\textwidth, clip, trim={0.5cm 0.5cm 0cm 0cm}]{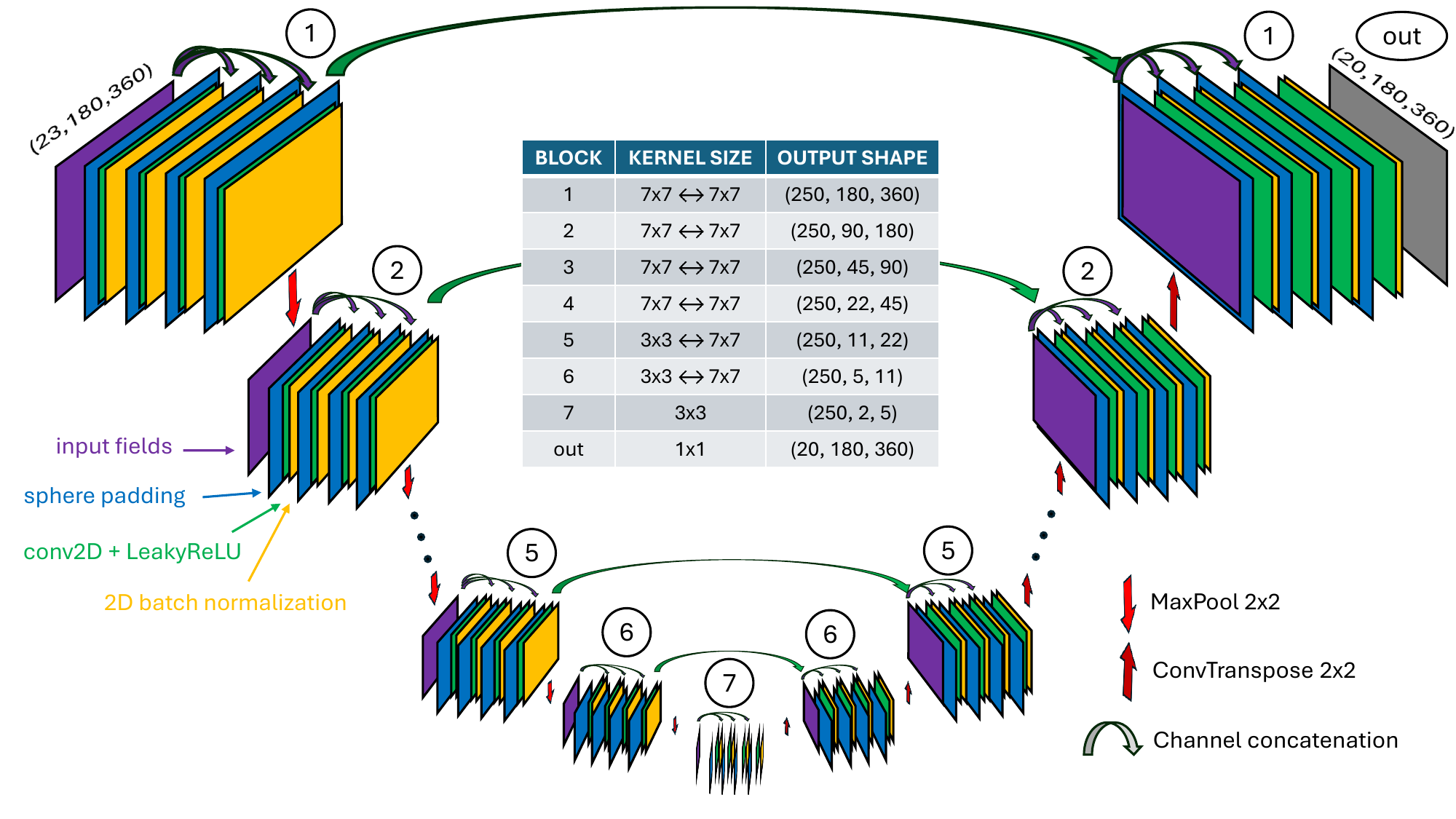}
    \caption{The structure of the neural-network forecasting model. The kernel sizes in the table correspond to the convolutional blocks in the encoding/decoding part of the network (left/right part of the scheme).}
    \label{fig:NNfwd}
\end{figure}

The forecasting model is a U-Net~\citep{Ronneberger2015UNet}, whose architecture is shown in Fig.~\ref{fig:NNfwd}. The network blocks have a similar design as in the AE, but the convolutional layers contain 250 channels. In the innermost layers in the downscaling part of the network, $3\times3$ kernels are used instead of the larger $7\times7$ ones. Additionally, blocks at the same resolution level are linked via skip connections, enabling the preservation and reuse of fine-scale features throughout the network. The NN was trained to produce 12-hour forecasts, while longer lead times were obtained by iteratively applying the model. A 24-hour forecast and its ERA5 reanalysis counterpart are compared in Fig.~\ref{fig:NNfwd quality}.

\section{Appendix B: 3D-Var cost function minimization algorithm}
\label{app:minimization}
\renewcommand{\thefigure}{B\arabic{figure}}
\setcounter{figure}{0}
The first guess for $\boldsymbol{\chi}$ in the minimization was a zero-vector, which corresponds to $\mathbf{z}=\mathbf{z}_b$. The gradient of the cost function was computed using automatic differentiation in Pytorch. The cost function was minimized using the stochastic gradient descent (SGD) optimizer with an initial step size of 0.3. The step size was halved, if $\mathcal{J}_\chi$ increased between iterations with a lower limit of $10^{-4}$. Minimization terminated once the gradient norm dropped below $1\,\%$ of its initial value, typically achieved in 6 to 12 steps. The maximum number of steps allowed was 50.

\newpage

\renewcommand{\thefigure}{S\arabic{figure}}
\setcounter{figure}{0}

\begin{landscape}
\section*{Suplementary material}

\vspace{1cm}
\begin{figure}[h!]
    \centering
    \includegraphics[width=\linewidth, clip, trim={0cm 0cm 2.5cm 0cm}]{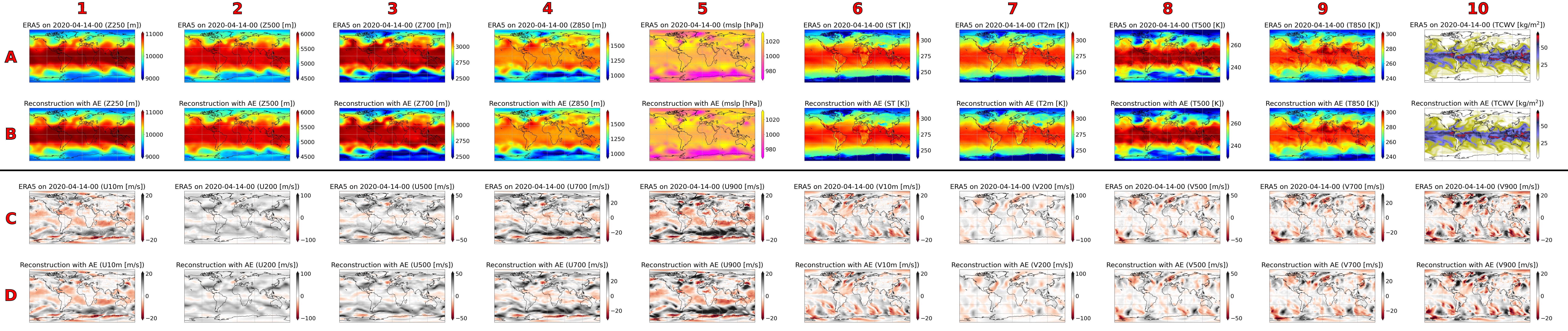}
    \caption{Comparison of the autoencoder (AE) outputs with their ERA5 targets. (Rows  A, C) ERA5 reanalysis on April 14, 2020, at 00 UTC. (B, D) Corresponding AE reconstructions of the fields shown in (A, C).}
    \label{fig:AE quality}
\end{figure}
\end{landscape}

\begin{landscape}
\begin{figure}[h!]
    \centering
    \includegraphics[width=\linewidth, clip, trim={0cm 0cm 2.5cm 0cm}]{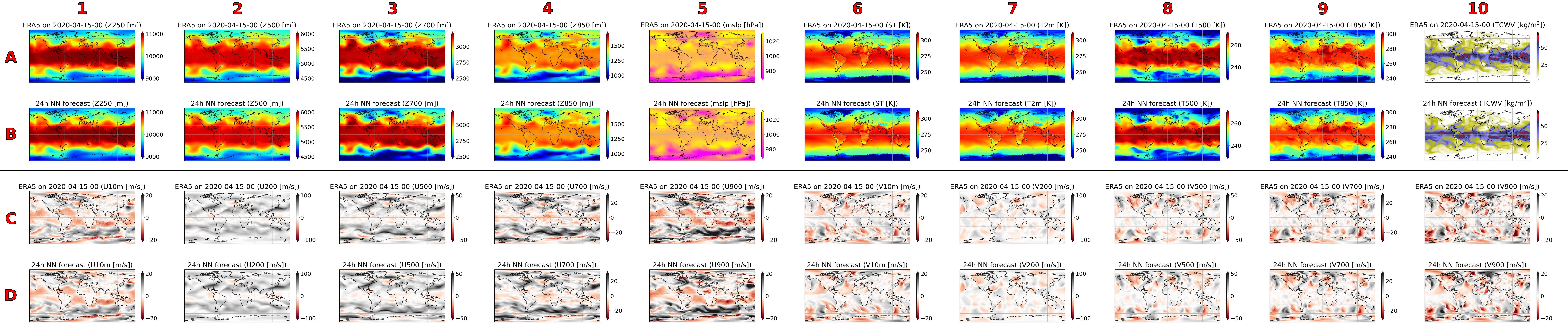}
    \caption{Comparison of the neural network (NN) forecasts with ERA5 reanalysis. (Rows A, C) ERA5 fields on April 15, 2020, at 00 UTC. (B, D) 24-hour forecasts from the NN model, valid at the same time.}
    \label{fig:NNfwd quality}
\end{figure}
\end{landscape}

\begin{figure}[h!]
    \centering
    \includegraphics[width=\textwidth]{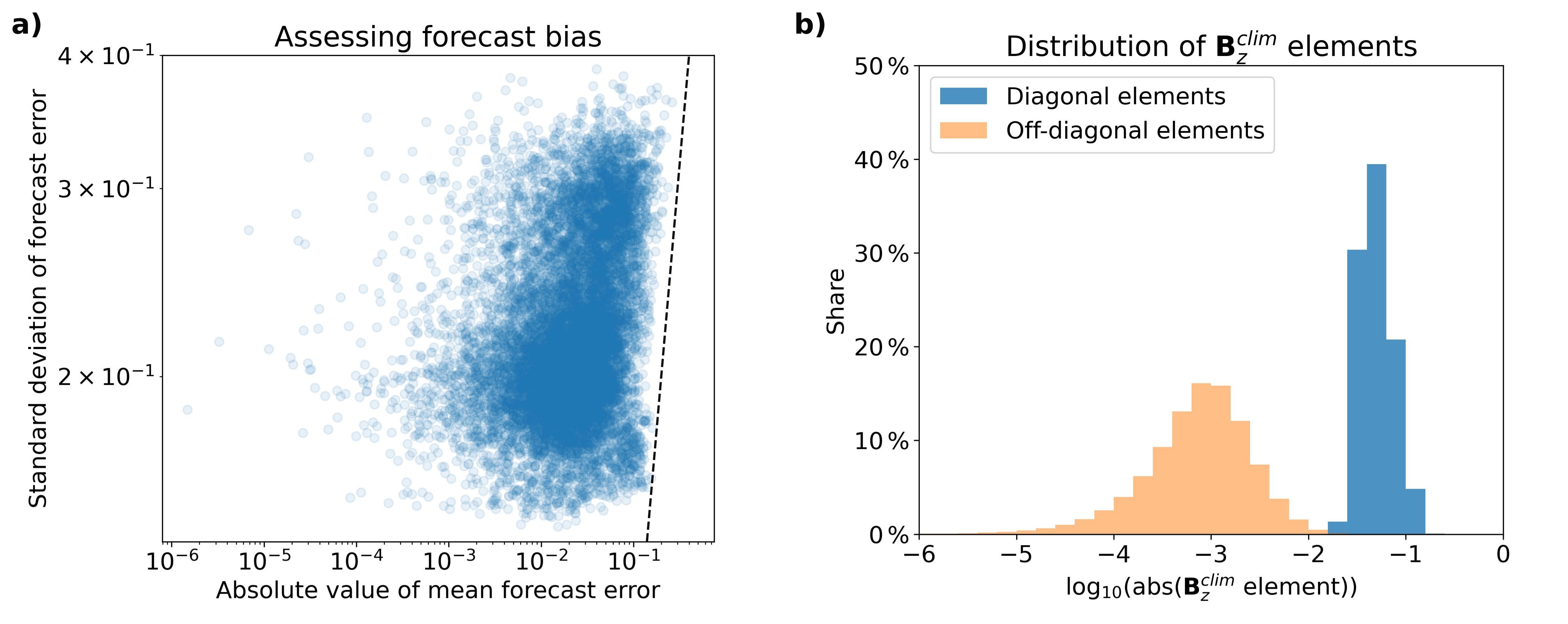}
    \caption{(a) Comparison of the forecast bias and forecast-error standard deviation for the 24-hour forecast used as background. Each point represents one of the 12100 latent vector elements, with absolute mean error on the x-axis, and error standard deviation on the y-axis, computed over the entire validation set. (b) Distribution of $\mathbf{B}_{z}^{clim}$ elements, shown using logarithmic bin width of 0.2.}
    \label{fig:B_and_bias}
\end{figure}

\begin{landscape}
\begin{figure}[h!]
    \centering
    \includegraphics[width=\linewidth, clip, trim={0cm 0cm 2.5cm 0cm}]{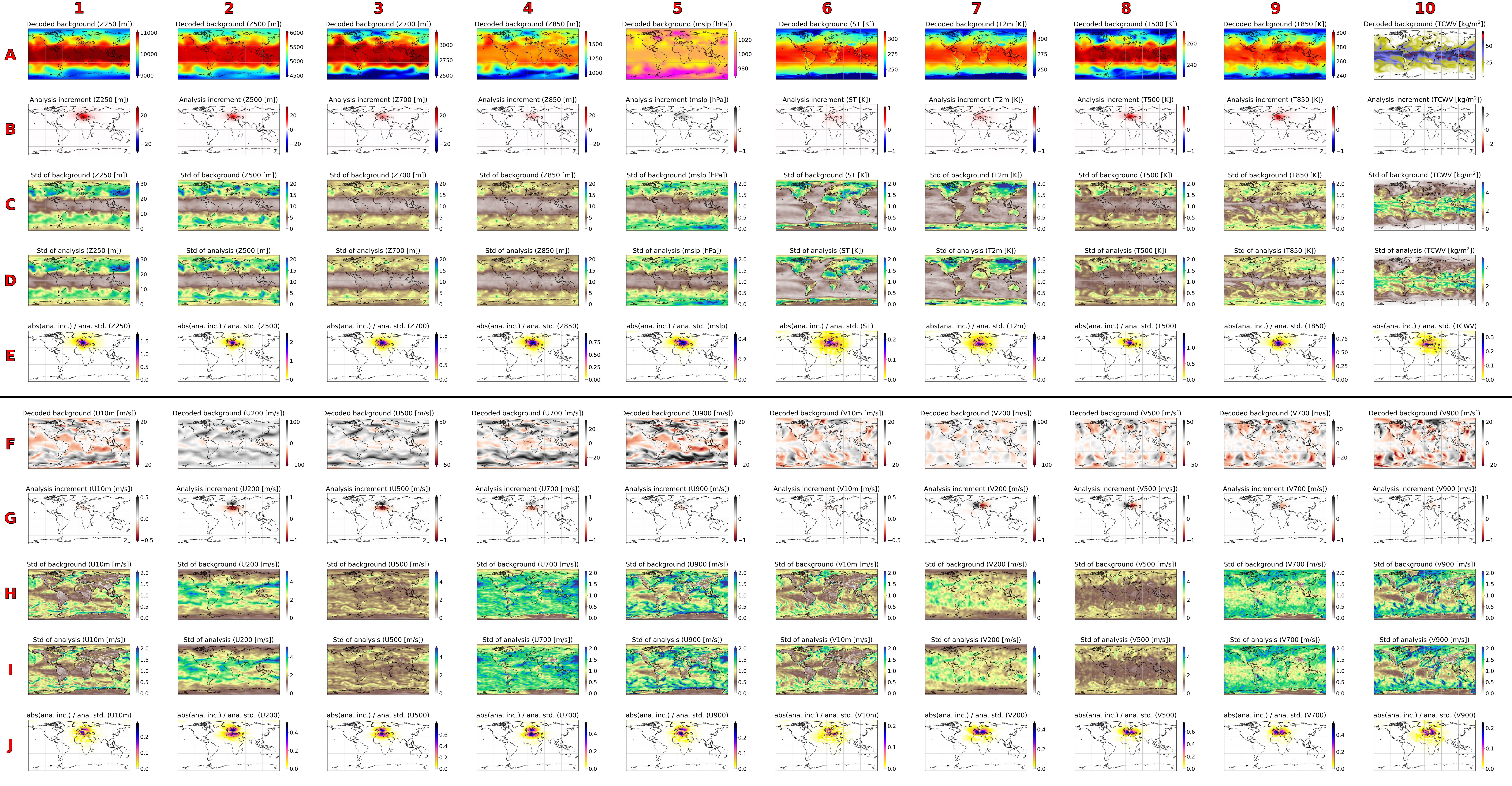}
    \caption{Single-observation DA experiment using a Z500 observation over Ljubljana, Slovenia, with a 30\,m departure and a 10\,m observation-error standard deviation. (Rows A and F) Decoded background fields for all 20 forecasted variables. (B and G) Analysis increments. (C and H) Background-error standard deviation. (D and I) Analysis-error standard deviation. (E and J) Ratio of the absolute value of the analysis increment and the analysis-error standard deviation.}
    \label{fig:full Ljubljana}
\end{figure}
\end{landscape}

\begin{figure}[h!]
    \centering
    \includegraphics[width=0.5\linewidth]{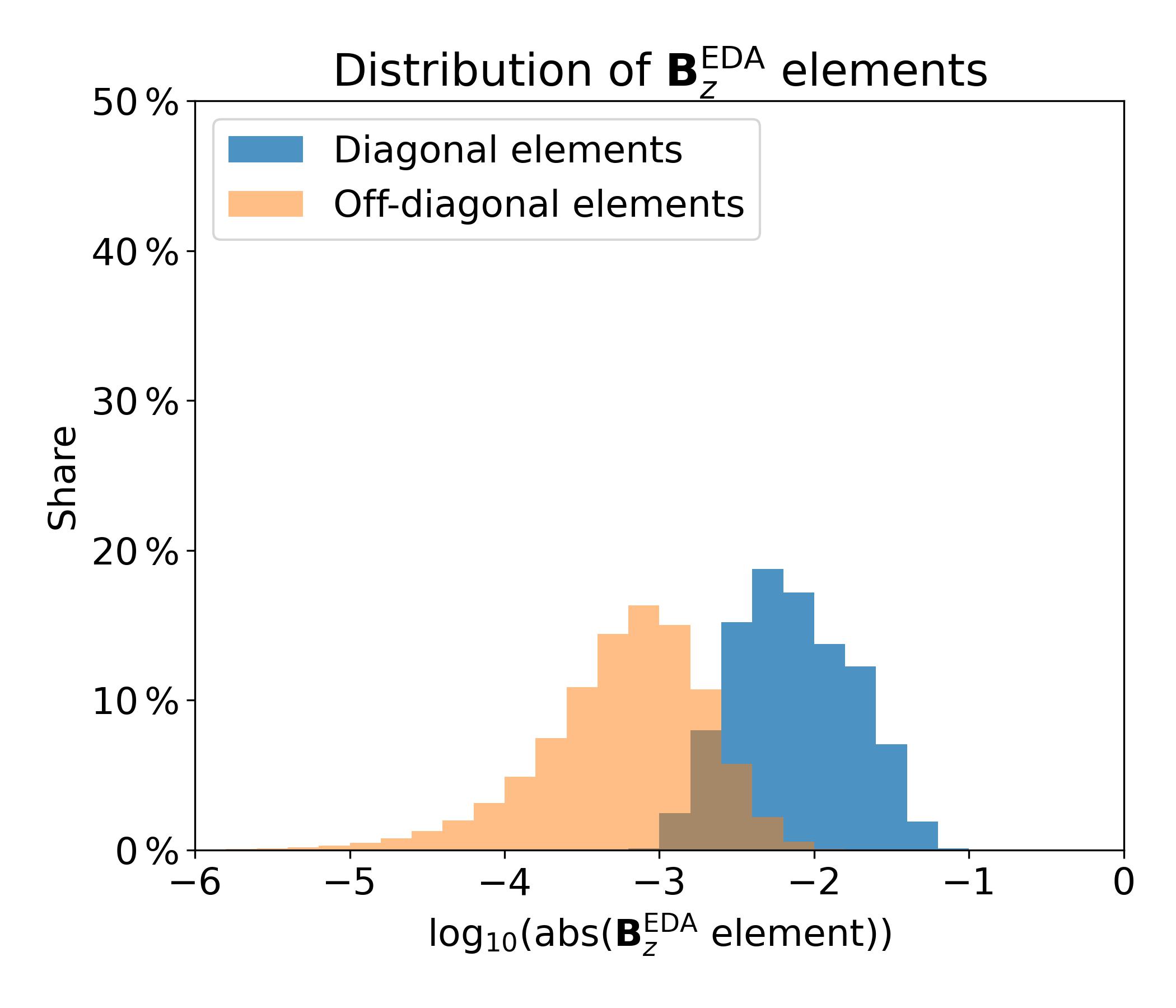}
    \caption{The distribution of $\mathbf{B}_{z}^\mathrm{EDA}$ elements. The bin width is 0.2 on the logarithmic scale.}
    \label{fig:B-matrix EDA}
\end{figure}

\begin{landscape}
\begin{figure}[h!]
    \centering
    \includegraphics[width=\linewidth, clip, trim={0cm 0cm 2.5cm 0cm}]{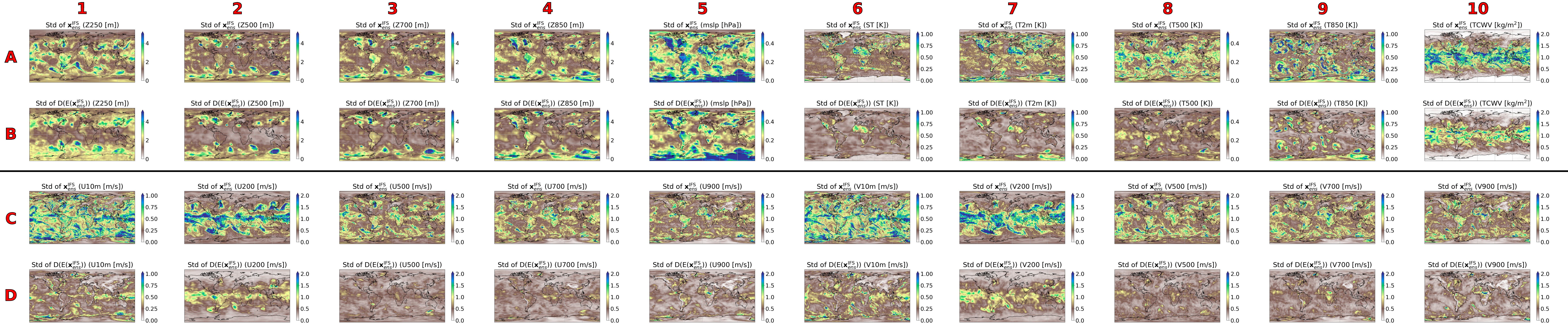}
    \caption{Standard deviation of all variables for the IFS ensemble before (rows A and C) and after (B and D) propagating it through the autoencoder.}
    \label{fig:full EDA std}
\end{figure}
\end{landscape}

\begin{landscape}
\begin{figure}[h!]
    \centering
    \includegraphics[width=\linewidth, clip, trim={0cm 0cm 2.5cm 0cm}]{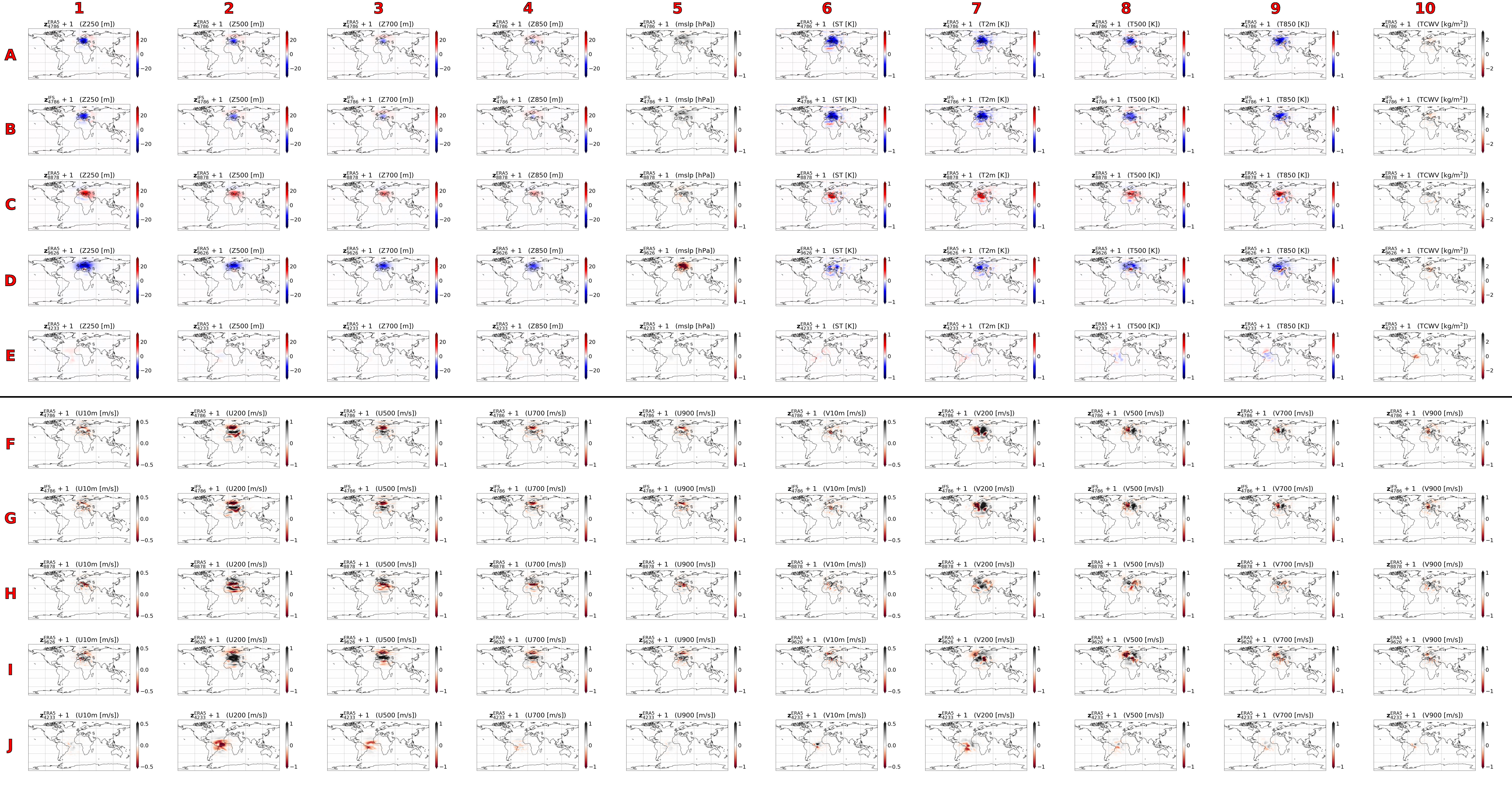}
    \caption{Changes of the decoded fields after modifying a single latent vector element. (Rows A and F) Changes after adding 1 to the latent vector element with index 4786 of the encoded ERA5 reanalysis for the 15th of April, 2020. (B and G) As (A and F), but for the encoded first IFS ensemble member for the 14th of April, 2024. (C and H) As (A and F), but for the latent vector element with index 8878. (D and I) As (A and F), but for the latent vector element with index 9626. (E and J) As (A and F), but for the latent vector element with index 4233. The examined elements are those that are highlighted in Fig.~\ref{fig:changes in latent space}.}
    \label{fig:basis functions}
\end{figure}
\end{landscape}

\end{document}